# Vibrational-mechanical properties of the highly-mismatched Cd$_{1-x}$Be$_x$Te semiconductor alloy – Experiment and *ab initio* calculations

A. Elmahjoubi,[1] M. B. Shoker,[1] O. Pagès,[1,*] V. J. B. Torres,[2] A. Polian,[3,4] A.V. Postnikov,[1] C. Bellin,[3] K. Béneut,[3] C. Gardiennet,[5] G. Kervern,[5] A. EnNaciri,[1] L. Broch,[1] R. Hajj Hussein,[1] J.-P. Itié,[4] L. Nataf,[4] S. Ravy,[4] P. Franchetti,[1] S. Diliberto,[6] S. Michel,[6] A. Abouais,[7,8] and K. Strzałkowski[7]

[1] Université de Lorraine, LCP-A2MC, ER 4632, F-57000 Metz, France

[2] Departamento de Fisica and i3N, Universidade de Aveiro, 3810 – 193 Aveiro, Portugal

[3] Institut de Minéralogie, de Physique des Matériaux et de Cosmochimie, Sorbonne Université — UMR CNRS 7590, F-75005 Paris, France

[4] Synchrotron SOLEIL, L'Orme des Merisiers Saint-Aubin, BP 48 F-91192 Gif-sur-Yvette Cedex, France

[5] Université de Lorraine, Laboratoire de Cristallographie, Résonance Magnétique et Modélisations, UMR 7036, Vandoeuvre-lès-Nancy, F-54506, France

[6] Université de Lorraine, CNRS, IJL, F-57000, Metz, France

[7] Institute of Physics, Faculty of Physics, Astronomy and Informatics, Nicolaus Copernicus University in Toruń, ul. Grudziądzka 5, 87-100 Toruń, Poland

[8] Chouaib Doukkali University of El Jadida, National School of Applied Sciences, Engineering Science for Energy Lab, El Jadida, Morocco

[*] Correspondence and requests for materials should be addressed to O.P.
(email: olivier.pages@univ-lorraine.fr)

The CdTe-BeTe semiconductor alloy that exhibits a dramatic contrast in bond covalency / stiffness presumably clarifying its vibrational-mechanical properties is used as a benchmark to test the limits of the percolation scheme worked out to explain the complex Raman spectra of the related but less contrasted Zn$_{1-x}$Be$_x$-chalcogenides. The test is done in the native zincblende phase of Cd$_{1-x}$Be$_x$Te by way of experiment ($x \leq 0.11$) – combining Raman scattering with X-ray diffraction at high pressure – and / or *ab initio* calculations ($x \sim 0 - 0.5$, $\sim 1$). The measured macroscopic bulk modulus ($B_0$) drops below the CdTe value on minor Be incorporation ($x$=0.11), at variance with a (quasi) linear $B_0$ vs. $x$ increase predicted by *ab initio* calculations ($x \leq 0.3$), thus hinting at large anharmonic effects in the real crystal. Yet, no anomaly occurs at the microscopic (bond) scale as the regular bimodal percolation-type Raman signal predicted *ab initio* for Be-Te at minor content ($x$~0, 0.5) can be (barely) detected in high-pressure / low-temperature experiment ($x$ =0.11). However, at large Be content ($x$~1) the *ab initio* bimodal Be-Te Raman signal "relaxes" all the way down to (a slight) inversion – giving rise to an unprecedented crossing of percolation sub-branches in a "Raman frequency vs. $x$" plot. The inverted doublet exhibits a remarkable intensity interplay under pressure, grasped within a linear dielectric approach. Altogether, the $B_0$-drop at $x$~0 and the inversion of the Raman signal at $x$~1 reflect a negative impact of the large bond mismatch on the $Cd_{1-x}Be_xTe$ mechanical-vibrational properties at various length scales and compositions. Yet, the percolation scheme applies as such to $Cd_{1-x}Be_xTe$ – albeit in a "relaxed" form, without further refinement. This enhances the scheme's validity as a generic descriptor of phonons in alloys.

Due to its direct optical band gap of 1.5 eV and high optical absorption coefficient, the cubic-zincblende II-VI CdTe semiconductor compound is almost ideal for solar energy conversion. The CdTe-based technology, stimulated by high conversion rates achieved over the past decade (exceeding 22% in laboratory) together with the stability of the photovoltaic devices under operating conditions, has now developed up to the industrial scale[1]. As early as in 1960's, the alloying of CdTe has been explored as a convenient means to finely tune its physical properties being of major interest as those of a semiconductor (the optical band gap $E_0$ and the lattice constant $a$), in view of targeted applications. The current study of the prospective $Cd_{1-x}Be_xTe$ II-VI semiconductor alloy is in this line – although it mainly concerns the mechanical-vibrational properties and not so much the optical ($E_0$) and structural ($a$) ones.

One conventionally refers to II-VI and III-V pseudobinary $A_{1-x}B_xC$ semiconductor alloys of first generation – still actively studied[2] – as those in which the A↔B substitution involves elements from the third to sixth rows of the periodic table. The A-C and B-C chemical bonds enter with physical properties (length $l$ and covalency $\alpha_c$; the values below – given cubed in the latter case – are cited from Refs.[3,4] unless specified) close within a few percent[4], which facilitates alloying. The CdTe-based $Cd_{1-x}Zn_xTe$, $CdSe_xTe_{1-x}$ and $CdS_xTe_{1-x}$ systems are like this, forming "well-matched" alloys (WMA's)[1]. In the exemplary III-V WMA, $Al_{1-x}Ga_xAs$ ($\frac{\Delta l}{l}$~8‰, $\frac{\Delta\alpha_c^3}{\alpha_c^3}$~5%), and the leading II-VI WMA, $Cd_{1-x}Hg_xTe$ ($\frac{\Delta l}{l}$~3‰, $\frac{\Delta\alpha_c^3}{\alpha_c^3}$~2.5%), the matching is nearly perfect[3].

In the 1990's, the emergence of semiconductor alloys involving second-row elements in substitution – like N, Be and O, – *e.g.*, N-dilute $GaAs_{1-x}N_x$ (Ref.[5]) and $Zn_{1-x}Be_x$-chalcogenides[6,7] – created a disruption. In fact, light elements with small covalent radii form extremely short bonds which, moreover, exhibit odd features of the $\alpha_c$ that governs the resistance of a bond to a distortion in shear (increasing with $\alpha_c$) and hence the stability of the lattice. So, with Be-bonding (BeS, BeSe, BeTe) the $\alpha_c$ achieves maximum among II-VI's, whereas, on the contrary, the $\alpha_c$ of N-bonded compounds (InN, GaN, AlN) hits a minimum among III-V's[4]. Consequently, in zincblende alloys such as $Zn_{1-x}Be_xSe$ and $GaAs_{1-x}N_x$ the bond properties do dramatically differ, by as much as ($\frac{\Delta l}{l}$~9%, $\frac{\Delta\alpha_c^3}{\alpha_c^3}$~33%) and (~20%, ~23%), respectively. The discrepancy softens as N is replaced by next-row P – to form $GaAs_{1-x}P_x$ (~3.5%, ~1%). The "bond mismatch" is likewise large for $ZnSe_{1-x}O_x$ (~21% – Ref.[8], ~16%) but small for $ZnSe_{1-x}S_x$ (~4.5%, ~6.8%). Hence, the second-generation alloys involving (Be, N, O) in substitution were singled out as forming a new class of "highly-mismatched alloys" (HMA's for short)[9].

HMA's – to which belongs $Cd_{1-x}Be_xTe$ studied in this work – attract attention because in certain cases their large bond mismatch dramatically impacts the electronic band structure. For instance, even slight N-incorporation into GaAs (anyway limited to a few percent[10]) induces a giant (negative) bowing of $E_0$ (of ~100 meV per atomic percent of N)[10], due to a band anticrossing which results from coupling of the extended states forming the host conduction band with a quasi-resonant highly localized impurity state[10-14]. Oxygen likewise induces an intermediate band within the bandgap of O-dilute $ZnTe_{1-x}O_x$[15,16] and $CdTe_{1-x}O_x$[17]. No similar outstanding features were detected with $Zn_{1-x}Be_x$-chalcogenides, which therefore received less attention so far. $E_0$ varies linearly with $x$ in $Zn_{1-x}Be_xTe$[18], or undergoes just a slight bowing in $Zn_{1-x}Be_xSe$[19]. More generally, the (quasi) linearity with composition governs all studied critical points of the electronic band structure of $Zn_{1-x}Be_x$-chalcogenides (considering the direct gaps between the upper valence band and the lower conduction band at the $\Gamma$, $L$ and $X$ points – namely $E_0$, $E_1$ and $E_2$, and the corresponding gaps involving the light hole valence band – notably $E_0 + \Delta_0$ and $E_1 + \Delta_1$, as assigned in Ref.[20]), for both $Zn_{1-x}Be_xTe$[21] and $Zn_{1-x}Be_xSe$[22].

However, the $Zn_{1-x}Be_x$-chalcogenides have held a central place in what regards vibrational properties. In early 2000's, pioneering Raman studies[23-25] revealed a bimodal signal per bond (1-bond→2-mode) that contested the classical view of a unique mode per bond (1-bond→1-mode) in random zincblende alloys[26]. This was explained in terms of sensitivity of the effective bond force constant ($k$) – probed by Raman scattering – to local Be- and Zn-like environments, casted into the percolation model (PM, Ref.[27] and refs. therein). Subsequently, the PM enabled a unified understanding, hitherto missing, of the Raman spectra of II-VI, III-V and IV WMA's with cubic (zincblende, diamond) and hexagonal (wurtzite) structures, suggesting the model's universality[27].

The bond force constant ($k$) and the bond length ($l$) are related via a basic rule that the former decreases as the latter is stretched, and vice versa. Referring this to the $Zn_{1-x}Be_x$-HMA's the variation in Be-related bond force constant (~10%) observed in Raman[25,27] and far-infrared absorption[28] experiments is correlated with a predicted variation in Be-related bond length (~2%, from *ab initio* calculations[29]) depending on whether the Be-related bond is in a Be-like or Zn-like environment. Unfortunately, this bimodal distribution of the Be-related bond length could not be resolved in extended X-ray absorption fine structure (EXAFS) measurements on a synchrotron[30,31], and is *a priori* hardly detectable via the pair distribution function in conventional total X-ray scattering experiments[32,33]. The same applies a fortiori to the less contrasted WMA's. For instance, the III-V (covalent) $Ga_{1-x}In_xAs$ and II-VI (ionic) $ZnSe_{1-x}Te_x$ alloys usually ranked among WMA's despite their significant lattice mismatch (~7%, Ref.[3]) likewise exhibit a unique bond length per species at a given $x$ value in EXAFS data[34-36]. Yet their Raman signal is bimodal per bond – of the percolation type[25,37]. Hence, EXAFS does not resolve the splitting of the A-C or B-C bond lengths of $A_{1-x}B_xC$ zincblende alloys – HMA's included – due to A-like or B-like local environments and thus yields a basic 1-bond→1-length description[38], whereas the Raman signal diversifies into a (percolation-type) 1-bond→2-mode pattern. In a nutshell, the lattice dynamics zooms deeper into the alloy disorder than the lattice relaxation. This is interesting on the fundamental side.

Besides, the PM opens applications. In particular, the phonons being sensitive to the local environment – as formalized within the PM, they can be used to elucidate whether the atom substitution is ideally random or not[27], the central issue when dealing with alloys. Apart from vibrational spectroscopies, the nuclear magnetic resonance (NMR) seems to be the only one in-house technique capable of addressing such issue quantitatively[39] – an example is given below. Alternatively, one may resort to EXAFS measurements (using a synchrotron) of the second-neigbor distances for the common-atom sublattice (*i.e.*, the Te-Cd-Te and Te-Be-Te ones for $Cd_{1-x}Be_xTe$) – which differ on account that each bond tends to preserve its natural length in an alloy[35,38].

In this work, the PM is tested on the emerging $Cd_{1-x}Be_xTe$ HMA that exhibits an even larger bond mismatch ($\frac{\Delta l}{l} \sim 15\%$, $\frac{\Delta \alpha_c^3}{\alpha_c^3} \sim 36\%$) than the $Zn_{1-x}Be_x$-chalcogenides. It is not *a priori* obvious whether the PM would apply to such special system. In the positive case, it will be instructive to elaborate on specific consequences from the PM, listed below from basic to more advanced ones[40]:

1. Only Be-Te should exhibit a distinct Raman doublet. Being small, Be has more room than Cd to move around in the Te-cage to accommodate the local strain created by the bond mismatch, with concomitant impact on the Raman frequencies being more diversified for Be-Te than for Cd-Te.

2. Out of the two sub-modes within the Be-Te doublet, the softer (harder) one would refer to Be (Cd)-like environments. Consider, *e.g.*, an isolated Cd atom embedded in BeTe. In its vicinity, the short Be-Te bonds suffer a compressive strain due to a competition with the longer Cd-Te ones within the given lattice spacing, hence vibrate at a higher frequency than the bulk-like Be-Te bonds away from Cd. An inverse argumentation applies to CdTe doped with Be.

3. The Be-Te doublet is expected to converge under pressure. This is due to a larger volume derivative of the bond ionicity $\partial f_i / \partial \ln V$ (with $f_i = 1 - \alpha_c$) for Be-Te (0.7453) than for Cd-Te (0.407)[3]. Hence, the Be-like environment hardens under pressure faster than the Cd-like environment does, so that thence stemming (softer) Be-Te Raman line within the percolation doublet drifts (upwards) under pressure faster. A classification of II-VI and III-V (WMA's and HMA's) alloys was accordingly suggested in Ref.[27].

4. This convergence either ends up into a phonon exceptional point at the crossing-resonance (scenario 1), or freely develops into an inverted doublet post resonance (scenario 2), depending on whether the (Be-Te) bond responsible for the vibration doublet is dispersed (self-connects in chains) or matrix-like (self-connects in bulk), respectively. The Raman intensity of the minor mode within the doublet undergoes, correspondingly, an extinction (scenario 1) or enhancement (scenario 2).

More generally, the ambition of this work is an integrated and coherent fundamental study of mechanical properties of the $Cd_{1-x}Be_xTe$ HMA at the macroscopic (via elasticity) and microscopic (sensitive to chemical bonds) scales, using $Zn_{1-x}Be_xTe$[40] as a suitable reference – based on proximity of Cd and Zn in the periodic table. Specifically, we combine high-pressure X-ray diffraction measurements on single crystals at the PSICHÉ and CRISTAL beamlines of SOLEIL synchrotron, in search for the macroscopic bulk modulus ($B_0$), with high-pressure Raman scattering measurements (on the same samples), probing the effective bond force constants in line with the above raised issues (1-to-4) around the PM. The discussion of $Cd_{1-x}Be_xTe$ experimental data is supported by high-pressure *ab initio* snapshots of the lattice relaxation, notably to determine the equation of state from which $B_0$ is issued, and of the lattice dynamics, with special attention to the Raman frequencies and intensities. Additional *ab initio* calculations are implemented to cover $x$ values beyond the experimental range (currently limited to 11 at.% Be). Various *ab initio* codes are used, *i.e.*, AIMPRO[41,42] (Ab Initio Modeling PROgram), SIESTA[43,44] and QE[45] (Quantum Expresso), depending on need – as specified in the course of the discussion.

Besides, we briefly test by ellipsometry and transmission if the $E_0$ vs. $x$ dependency is (quasi)linear, like with $Zn_{1-x}Be_x$-HMA's, or significantly deviates from linearity, like with N/O-dilute HMA's. It is a matter to appreciate on an experimental basis whether HMAlloying with second-row elements (Be, N, O) is virtuous for $E_0$ (in that it generates a smooth linear-like $E_0$ vs. $x$ variation) only in case of Zn↔Be substitution (as discussed above) – for whatever reason, or is a more general rule with Be.

## Results and discussion

The studied samples consist of high-quality / purity $Cd_{1-x}Be_xTe$ and $Zn_{1-x}Be_xTe$ bulk single crystals grown by the Bridgman method with small Be content ($x$ ≤0.11 and 0.21, as determined by chemical analysis – see methods – and via the $a$ vs. $x$ linearity[46], correspondingly). The zincblende structure, common to end compounds, was confirmed by powder X-ray diffraction (Figs. S1a and S2; "S" stands for Supplementary Information) throughout intermediate compositions. The $Cd_{1-x}Be_xTe$ lattice expands / shrinks homothetically as $x$ varies (reflected by a linear $a$ vs. $x$ variation between the CdTe – this work – and BeTe[47] values, Fig. S1b), that seems to be common in alloys, HMA's included[2]. Further structural insight at the microscopic scale is gained via powder ($x$=0.07) $^{125}$Te solid-state nuclear magnetic resonance (NMR) measurements[39]. A bimodal NMR pattern (Fig. 1a) distinguishes between two tetrahedral environments for Te among five possible ones depending on the number of Cd and Be nearest neighbors. The NMR peak intensities scale as the fractions of tetrahedral Te-clusters with 4×Cd and (3×Cd, 1×Be) atoms at the vertices, as suggested by the Bernoulli's binomial distribution[39] at $x$=0.07 (Fig. 1a, inset). One notes a practical absence of Te-centered clusters with more Be atoms, consistently with experiment. Altogether the NMR data point towards an ideally random Cd↔Be substitution in $Cd_{0.93}Be_{0.07}Te$, presumably valid in all studied samples owing to the close Be contents.

Hence the discussed experimental trends hereafter are presumably intrinsic to random $Cd_{1-x}Be_xTe$-alloying. Additional ($^9$Be and $^{117}$Cd) NMR experiment completing the current ($^{125}$Te) NMR insight into $Cd_{0.93}Be_{0.07}Te$ are reported as Supplementary Information (Fig. S4).

The Tauc plots of $Cd_{1-x}Be_xTe$ transmission data ($x \leq 0.11$, Fig. S5a) reveal a quasi linear $E_0$ vs. $x$ trend between the CdTe[48,49] and BeTe[20] values (Fig. 1b). The linearity persists with $E_1$ and $E_1 + \Delta_1$ (Fig. 1b) accessed by ellipsometry (Fig. S5b) – and also with $E_2$ (not shown) notwithstanding a poor signal-to-noise ratio due to the lack of luminous flux at high energy with the light source used. This resembles the case of $Zn_{1-x}Be_x$-chalcogenides[18-22], contrasting that of dilute nitrides/oxydes[10-17]. Hence, the $E_0$ vs. $x$ linearity seems to be the rule with Be substitution. The (Be, Zn, Cd) substituents involved in $(Zn,Cd)_{1-x}Be_x$-chalcogenides are nearly iso-electronegative (within few percent), in contrast with dilute nitrides / oxydes that exhibit a large contrast in electronegativity between alloying elements (in the range 25-40%) – the admitted cause[16] for their large (negative) $E_0$ vs. $x$ bowing[12-16]. This might be the reason why the $(Zn,Cd)_{1-x}Be_x$-HMA's behave like WMA's in what regards their optical properties.

We turn now to mechanical-vibrational properties, independently probed at the macroscopic scale at SOLEIL synchrotron (PSICHÉ beamline), searching for the bulk modulus $B_0$ by high-pressure X-ray diffraction (HP-XRD), and at the microscopic scale by high-pressure Raman scattering (HP-RS), sensitive to the bond force constants, in relation to the raised issues (1-to-4) around the PM. For the sake of consistency, both studies are performed at the same Be content ($x$=0.11).

The sequence of pressure-induced structural transitions (Fig. 1c) apparent in the $Cd_{0.89}Be_{0.11}Te$ HP-XRD diffractograms (taken on the upstroke, Fig. S1c) repeats that of pure CdTe (0 GPa: zincblende/ZB-cubic; 4 GPa: Rocksalt/RS-cubic, 12 GPa: Cmcm-orthorhombic)[50], however shifted to higher pressures (ZB→RS: 5 GPa, RS→Cmcm: 16 GPa) due to a reinforcement of the soft (ionic) CdTe-lattice by the stiff (covalent) Be-bonding – coming about also in $Zn_{1-x}Be_xSe$[30] and $Zn_{1-x}Be_xTe$[40]. One disconcerting feature, though, is that $B_0$ derived for $Cd_{0.89}Be_{0.11}Te$ (41.667 $\pm$ 0.243 GPa – the error bar is within the symbol size, Fig. 1d) from best fitting of the experimental pressure vs. volume dependence via the Birch-Murnaghan equation[51] (Fig. S1d) drops below the $B_0$ value of CdTe[52] (43.7 $\pm$ 1.0 GPa, Fig. 1d, symbol) – which in turn is well below the BeTe one[53] (~67 GPa).

The $B_0$-drop is not accidental, due to the quality of fit. The first-order pressure derivative of $B_0$ evaluated at 0 GPa, $B'_0$, coming out in the fit is 4.00 $\pm$ 0.15 for $Cd_{0.89}Be_{0.11}Te$, nearly matching the CdTe[51] (3.8 $\pm$ 0.6) and BeTe (fixed to 4 in Ref.[53]) values. By adopting for $B'_0$ the "optimized" AIMPRO ($B'_0$=4.7, $B_0$=40.5 GPa) and SIESTA ($B'_0$=5.1, $B_0$=39.7 GPa) values at ~10 at.% Be (see below), the $B_0$-drop for $Cd_{0.89}Be_{0.11}Te$ is emphasized (and the quality of fit degraded – not shown). A similar $B_0$-drop is also evidenced by HP-XRD diffraction with $Zn_{1-x}Be_xTe$ (Fig. S3b, x≤0.21) – using $B'_0$=4 (found relevant for both ZnTe[54] and BeTe[53] parents in experiment) for the fitting (Fig. S3a), hence apparently a common feature of BeTe-based alloys.

The $B_0$-drop deviates from the (quasi) linear $B_0$ vs. $x$ experimental trend observed with $Zn_{1-x}Be_xSe$ – however, disrupted by a punctual lattice hardening on percolation of the stiff Be-Se bonds[30]. The reason why $Cd_{1-x}Be_xTe$ and $Zn_{1-x}Be_xTe$, but not $Zn_{1-x}Be_xSe$, suffer a $B_0$-drop from minor Be incorporation, might relate to a much weaker ionicity of the Be-Te chemical bonding ($f_i$=0.222) compared with Be-Se ($f_i$=0.420)[3]. This gives rise to more dramatic $\Delta f_i/f_i$-contrasts in the BeTe-based $Cd_{1-x}Be_xTe$ (~70%) and $Zn_{1-x}Be_xTe$ (~60%) alloys than in the BeSe-based $Zn_{1-x}Be_xSe$ one (~43%)[3], resulting in more severe bond distortions eventually affecting the mechanical properties in bulk.

Remarkably, no such $B_0$-drop but a linear $B_0$ vs. $x$ trend is suggested by *ab initio* results ($x \leq 0.3$, hollow symbols, Fig. 1d). The linearity seems robust since it is verified by two independent sets of calculations, applying the SIESTA (64-atom supercells, squares) and AIMPRO (for 216-atom supercells, diamonds) codes to distinct sets of fully- and partially-relaxed quasirandom (see methods) pseudobinary atomic arrangements, correspondingly. The *ab initio* vs. experiment discrepancy around $B_0$ suggests large anharmonic effects in real $Cd_{1-x}Be_xTe$, developing on large length scales – involving the full variety of atom arrangements making up a real randomly-disordered alloy, *i.e.*, beyond the finite supercell-sizes used for the current *ab initio* calculations.

Now we address microscopic / vibrational properties of $Cd_{1-x}Be_xTe$, in line with the PM-issues around the Be-Te Raman signal (1-to-4) raised in the introductory Sec. A comparison with the reference Be-Te Raman percolation doublet of $Zn_{0.89}Be_{0.11}Te$ – recently studied in detail[40] – matched in Be content with $Cd_{0.89}Be_{0.11}Te$, helps to fix ideas. Two impurity modes are involved in this case, a lower / minor $TO_{Be-Te}^{Be}$ and an upper / dominant $TO_{Be-Te}^{Zn}$ – with subscript and superscript referring to the vibrating bond and to the local environment, respectively – separated by $\Delta_{Be-Te}$~20 cm$^{-1}$, the intensities of these modes differing by roughly an order of magnitude.

Fig. 2a displays an overview of the Cd-Te and Be-Te $Cd_{1-x}Be_xTe$ TO Raman signals across the composition domain with information on frequency (curves) and intensity (color code) – technical detail is given below – offering a convenient eye-support for the forecoming discussion of the $x$-dependence of the Be-Te Raman doublet of $Cd_{1-x}Be_xTe$. Experimental TO and LO Raman frequencies of pure BeTe taken from the literature[55] are added, for reference purpose. The given overview unveils in advance the main result of this work, namely, an irregular crossing of the two Be-Te "percolation" sub-branches ($x$~0.8) contrasting with the regular parallelism of the Cd-Te ones.

Raman measurements at 0 GPa / 300 K with the red (632.8 nm) and blue (488.0 nm) laser lines (nearly resonant with the $E_0$ and / or $E_0 + \Delta_0$ electronic transitions of Be-dilute $Cd_{1-x}Be_xTe$, Fig. 1b) fail to reveal the Be-Te doublet (Fig. S7). Only one Be-Te mode is visible, at ~390 cm$^{-1}$, consistent with the scarce experimental (far-infrared) data in the literature[56] and existing calculations using the Green's function theory[57], far away from the (TO, LO) CdTe-lattice band (140 – 170 cm$^{-1}$). This line is assigned as the upper / main $TO_{Be-Te}^{Cd}$, by analogy with $TO_{Be-Te}^{Zn}$ of Zn$_{1-x}$Be$_x$Te. The lower / minor $TO_{Be-Te}^{Be}$ band is not visible, presumably screened by the second-order CdTe-lattice signal ($2LO_{Cd-Te}$) that emerges as a strong feature nearby, being emphasized / reduced with the red / blue laser excitation.

The conditions for testing the PM are improved by keeping the blue laser line but working at high pressure and low temperature (80 K), that offers a number of benefits. First, the Be-Te signal sharpens due to the increased phonon lifetime. Second, the pressure domain of the native zincblende phase enlarges[58]. Third, the low temperature slows down the formation of Te aggregates under intense laser exposition such as achieved by focusing the laser beam onto a tiny sample placed in a diamond anvil cell – a notorious problem with CdTe-like crystals[59]. The Raman spectra taken in the upstroke up to 4.3 GPa in the native zincblende phase of Cd$_{0.89}$Be$_{0.11}$Te (Fig. 2b) transiently reveal the target lower / minor $TO_{Be-Te}^{Be}$ (marked by an asterisk) on the low-frequency side of the main $TO_{Be-Te}^{Cd}$ at ~3.5 GPa, before its partial resorption at ~4.3 GPa. This is consistent with $TO_{Be-Te}^{Be}$ suffering a progressive collapse while converging towards $TO_{Be-Te}^{Cd}$ prior to its extinction at the crossing / resonance, as observed with the reference Be-Te doublet of Zn$_{0.89}$Be$_{0.11}$Te across a similar pressure range (2.3-7.9 GPa)[40]. This fits into scenario 1 of the convergence process (cf. the issue 4).

Although the analogy with Zn$_{89}$Be$_{11}$Te is enlightening, it remains limited to grasp the behaviour of Cd$_{0.89}$Be$_{0.11}$Te. Additional support is searched for by calculating the high-pressure *ab initio* (AIMPRO) Cd$_{1-x}$Be$_x$Te Raman spectra at small-to-moderate ($x$~0, 0.5) Be content. However, a limit to pressure is set by the supercells becoming unstable from 10 GPa ($x$~0, not shown) and 15 GPa ($x$~0.5, Fig. S7) onwards. The stability improves at large Be content ($x$~1), also considered to complete an *ab initio* Raman insight at well-spanned $x$ values across the composition domain (Figs. 2c, 2d and 2e).

Paired impurities (forming a duo connected via Te, as schematized in Fig. 2a) in large parent supercells represent the minimal impurity motif offering a clear distinction, in the context of the PM, between "impurity" and "host" vibrations in "domestic" and "foreign" environments (labels $i$-to-$iv$ in Fig. 2a), hence four situations in total at both ends of the composition domain ($x$~0 – Fig. 2c and $x$~1 – Fig. 2e). The duo-impurity modes are identified via their wavevectors, depending on whether they point along the duo (in-chain, noted $ii$ and symbolized by ↔) or perpendicular to it (out-of-chain, noted $i$ and marked ↮, yielding five possible variants[40]). For the host / dominant bond species, the distinction between "foreign" and "domestic" environments is established through the Raman intensity, being small next to the impurity-duo (close-duo, noted $iv$, the "foreign" case) and large away from it (bulk, noted $iii$, the "domestic" case). Additional insight at maximum alloy disorder ($x$=0.5 – Fig. 2d) using a nominally random Cd$_{54}$Be$_{54}$Te$_{108}$ supercell (see methods) completes the *ab initio* picture.

*Ab initio* results do reveal an actual Cd-Te percolation doublet for Cd$_{1-x}$Be$_x$Te at $x\sim0$ (Fig. S6a) and 1 (Fig. S6b), albeit a compact one (Fig. 2a) – cf. the issue 1, with $TO_{Cd-Te}^{Be}$ set below $TO_{Cd-Te}^{Cd}$ – cf. the issue 2, as expected. The frequency gap $\Delta_{Cd-Te}$ hardly exceeds a few cm$^{-1}$ (that won't be detectable in experiment) meaning that the Cd-Te vibration is almost blind to the local environment. At $x\sim1$ the upper $TO_{Cd-Te}^{Cd}$ is frozen (not Raman active, the frozen mode – identified by its wavevector – is spotted by an arrow in Fig. S6b) due to a phonon exceptional point being achieved already at 0 GPa – cf. the issue 4. At $x\sim0$ the Cd-Te doublet becomes inverted by increasing pressure to 5 GPa (Fig. S6a) – cf. the issues 3 and 4. Altogether, this offers a perfect analogy with the Zn-Te doublet of the reference Zn$_{1-x}$Be$_x$Te case[40].

The analogy between Cd$_{1-x}$Be$_x$Te and Zn$_{1-x}$Be$_x$Te is not as clear in their common Be-Te spectral range, successively addressed hereafter with Cd$_{1-x}$Be$_x$Te at small, intermediary and large Be contents.

At $x\sim0$ (Fig. 2c), the Be-duo generates a nominal percolation-type (lower / in-chain $TO_{Be-Te}^{Be}$, upper / out-of-chain $TO_{Be-Te}^{Cd}$) Raman doublet at 0 GPa with a large separation ($\Delta_{Be-Te}\sim35$ cm$^{-1}$) – conforming to the issue 2. Not surprisingly, the corresponding ($TO_{Be-Te}^{Be}$, $TO_{Be-Te}^{Zn}$) doublet in Zn$_{1-x}$Be$_x$Te is less resolved ($\Delta_{Be-Te}\sim20$ cm$^{-1}$)[40] due to the smaller bond ($\frac{\Delta l}{l}$, $\frac{\Delta\alpha_c^3}{\alpha_c^3}$)-contrast in Zn$_{1-x}$Be$_x$Te ($\sim9\%, \sim31\%$) than in Cd$_{1-x}$Be$_x$Te ($\sim15\%, \sim36\%$). As pressure is increased $TO_{Be-Te}^{Be}$ gets closer to $TO_{Be-Te}^{Cd}$ – cf. the issue 3 – and suffers a major collapse, to such extent that both its Raman intensity and $\Delta_{Be-Te}$ are halved (Fig. 2c), *i.e.*, the early signs of a phonon exceptional point on the verge of being achieved, in line with scenario 1 – cf. the issue 4. Generally, this finds echo in experimental findings – even though $TO_{Be-Te}^{Be}$ shows up as a mere shoulder on $TO_{Be-Te}^{Cd}$ in experiment (marked by an asterisk in Fig. 2b), whereas it emerges as a distinct feature in *ab initio* data (Fig. 2c). This might relate to a basic difference at minor Be content that *ab initio* calculations run on an "ideal" lattice – verified by the $B_0$ vs. $x$ linearity – whereas the real lattice suffers a massive $B_0$-drop in experiment (Fig. 1d).

At $x\sim0.5$ (Fig. 2d), scenario 1 still applies, as expected – cf. the issue 4. At 0 GPa, the Be-Te Raman signal of Cd$_{54}$Be$_{54}$Te$_{108}$ ($x$=0.5) shows up as a unique broad $TO_{Be-Te}$ band. At 5 GPa, this latter transforms into a compact percolation-type ($TO_{Be-Te}^{Be}$, $TO_{Be-Te}^{Cd}$) doublet subsided on its low-frequency side. At 10 GPa, the doublet further shrinks and the subsidence is emphasized to such extent that, apparently, only $TO_{Be-Te}^{Cd}$ survives, and $TO_{Be-Te}^{Be}$ is killed. This nicely recapitulates the sequence leading to the achievement of a phonon exceptional point – cf. the issue 4. The sequence is interrupted from 15 GPa on, due to the collapse of the zincblende structure in the supercell used, manifested by a departure of bond angles from the nominal tetrahedral value (109°, Fig. S7).

At $x\sim1$ (Fig. 2e), the situation becomes irregular. While well separated when stemming from the Be-duo ($x\sim0$), the ($TO_{Be-Te}^{Be}$, $TO_{Be-Te}^{Cd}$) doublet of Cd$_{1-x}$Be$_x$Te becomes quasi degenerate ($TO_{Be-Te}^{Be}\equiv TO_{Be-Te}^{Cd}$) at 0 GPa when due to the Cd-duo ($x\sim1$). Indeed, various minor features composing $TO_{Be-Te}^{Cd}$ (marked by asterisks in Fig. 2e) do overlap with the main $TO_{Be-Te}^{Be}$. In contrast, the ($T_{Be-Te}^{Be}$, $TO_{Be-Te}^{Zn}$) doublet of the reference $Zn_{1-x}Be_xTe$ case is globally preserved across the composition domain[40].

The quasi degeneracy of the Be-Te Raman doublet of $Cd_{1-x}Be_xTe$ at $x\sim1$ from 0 GPa undermines the possibility for scenario 2 to develop under pressure – cf. the issue 4 – offering a novel case study.

Under pressure, the $TO_{Be-Te}^{Be}$ and $TO_{Be-Te}^{Cd}$ lines, already close at 0 GPa, are forced to further proximity – because the local environment becomes less discriminatory between like bond vibrations under pressure[40] – and hence do couple mechanically. This mediates a transfer of the oscillator strength from the main to the minor mode, directly impacting the Raman intensities. Remarkably, the Be-Te oscillator strength is channeled from high to low frequency in the $Cd_{1-x}Be_xTe$ case ($x\sim1$, as emphasized by a vertical arrow in Fig. 2e), and not the other way around as observed in the reference $Zn_{1-x}Be_xTe$ ($x\sim1$) case[40]. This can be explained only if the minor $TO_{Be-Te}^{Cd}$ (recipient of oscillator strength) lies below the main $TO_{Be-Te}^{Be}$ (donor of oscillator strength) at 0 GPa. Hence, at 0 GPa the Be-Te doublet of $Cd_{1-x}Be_xTe$ ($x\sim1$) is inverted relative to its regular $Zn_{1-x}Be_xTe$ counterpart ($x\sim1$) – and not only degenerated. Specifically, the vibration lines follow (upwards) in the ($TO_{Be-Te}^{Cd}$, $TO_{Be-Te}^{Be}$) order – pay attention to the superscripts, as opposed to the regular ($TO_{Be-Te}^{Be}$, $TO_{Be-Te}^{Zn}$) order with $Zn_{1-x}Be_xTe$[40] ($x\sim1$) – cf. the issue 2.

In earlier work[60] we have shown that the shape of the upper (best-resolved) TO percolation doublet in a "Raman frequency vs. $x$" plot may vary a lot (parallel branches vs. trapezoidal / triangular distortions) depending on wether the parent TO is dispersive or not. The main arguments are briefly recalled hereafter by focusing on the lower parent–$iv$ and upper impurity–$iii$ modes forming the percolation doublet in the parent limit. Only the latter mode is subject to dispersion, not the former one, so that their comparison provides a straightfoward insight into the dispersion effect.

In absence of dispersion (as, e.g., for GaP), the $iii-iv$ frequency gap is governed by the local strain due to the bond mismatch (between, e.g., Ga-As and Ga-P in the case of $GaAs_{1-x}P_x$), resulting in parallel branches across the composition domain[60]. The TO mode of CdTe is nearly dispersionless[61], so that the Cd-Te doublet of $Cd_{1-x}Be_xTe$ actually exhibits such parallelism (Fig. 2a). In case of a positive (resp. negative) dispersion, $iii$ is upward (resp. downward) shifted with respect to $iv$ roughly by the magnitude of the dispersion, with concomitant on the $iii$–$iv$ frequency gap being enlarged – trapezoidal distortion – (resp. squeezed – triangular distortion), as observed[60] with $ZnSe_{1-x}S_x$ (resp. $Si_{1-x}Ge_x$).

We are mainly interested in the negative TO dispersion (with $Cd_{1-x}Be_xTe$, see below). In this case an inversion of the percolation doublet occurs if the dispersion effect outweights the effect of the local strain (as observed, e.g., with the Si-Si doublet of $Si_{1-x}Ge_x$ at $x\sim0$ – Ref.[60]). Such an inversion does not occur in the dilute limit, because, being impurity modes, the $i$ and $ii$ forming the percolation doublet at this limit suffer a similar dispersion effect. Since the percolation doublet is inverted in the parent limit but regular in the dilute one, its two sub-branches should cross at a certain composition. Such crossing was not observed yet – not even with $Si_{1-x}Ge_x$ (due to the $i$-$ii$ degeneracy[60] at $x$~1).

The TO mode of BeTe exhibits a large negative dispersion[62] (~50 cm$^{-1}$). Accordingly, in both $Zn_{1-x}Be_xTe$[40] and $Cd_{1-x}Be_xTe$ the Be-Te splitting is smaller in the parent limit ($x$~1, $iii$-$iv$~7 cm$^{-1}$ – Ref.[40] and ~0 cm$^{-1}$ – Fig. 2d, respectively) than in the dilute one ($x$~0, $iii$-$iv$~30 cm$^{-1}$ – Ref.[60] and ~45 cm$^{-1}$ – Fig. 2c, respectively). Based on *ab initio* (SIESTA) calculations done on the Be-Te impurity—$i$ mode of $Cd_{1-x}Be_xTe$ (Sec. SII.1)– taken as representative for all ($i$-to-$iii$) Be-Te impurity modes – the Be-Te doublet at $x$~0 ($i$-$ii$) is downshifted by ~11 cm$^{-1}$ due to the dispersion effect (short vertical arrows in Fig. 2a at $x$~0). A crude view of the (virtual) Be-Te doublet of $Cd_{1-x}Be_xTe$ due to the sole effect of the local strain, *i.e.*, deprived of dispersion, is reconstructed by drawing parallel Be-Te sub-branches (dashed lines, taken straight in a first approximation) – by analogy with the dispersionless Ga-P doublet of $GaAs_{1-x}P_x$ (Ref.[60]) – between the (virtual) Be-Te doublet inferred at $x$~0 in absence of dispersion and the same doublet attached to the parent TO mode at $x$~1.

When confronted with *ab initio* data at $x$~1 – taking into account blindly the effects of strain and dispersion, the above-derived (virtual) strain-related Be-Te doublet appears to be seriously challenged by the TO dispersion, whether considering $Cd_{1-x}Be_xTe$ or $Zn_{1-x}Be_xTe$. In the latter case, however, the effect of the local strain dominates so that the ordering of Be-Te branches is nicely preserved across the composition domain[40], even if the branches do not run exactly parallel (cf. the comparison between the $i$-$ii$ and $iii$-$iv$ frequency gaps). In $Cd_{1-x}Be_xTe$ the effect of the dispersion achieves maximum at $x$~1 (long vertical arrow, Fig. 2a) and overwhelms the effect of the local strain. This leads to an inversion of the Be-Te doublet ($x$~1), preceded by the crossing of Be-Te sub-branches ($x$~0.8) – an unprecedented case among all re-examined alloys within the PM so far.

Generally, the small / large Be-Te dispersion effect in $Zn_{1-x}Be_xTe$ / $Cd_{1-x}Be_xTe$ at $x$~1 reflects a small / large (negative) impact of the local lattice distortions due to the Zn / Cd ↔ Be atom substitution on the Be-Te vibrations (the minimum / maximum Be-Te dispersion effect at $x$~0 / ~1 in $Cd_{1-x}Be_xTe$ – compare the lengths of arrows in Fig. 2a – can be discussed on the same line). However, $Cd_{1-x}Be_xTe$ is not so much remarkable than $Zn_{1-x}Be_xTe$ with this respect, as the dispersion effect usually achieves maximum for the impurity modes of alloys, even in case of WMA's. The distorted percolation doublets of $ZnSe_{1-x}S_x$ and $Si_{1-x}Ge_x$ were discussed on this basis[60]. The Ga-As impurity mode likewise experiences the same (maximal) dispersion effect whether taken in the lattice-matched $Ga_{1-x}Al_xAs$ ($\frac{\Delta l}{l}$~8‰) or lattice-mismatched $Ga_{1-x}In_xAs$ ($\frac{\Delta l}{l}$~7%) alloys[63].

At this stage, the inversion of the Be-Te Raman doublet at $x \geq 0.8$ is (qualitatively) explained. Now we examine how the mechanical coupling develops between the inverted Be-Te submodes when forced to (further) proximity by pressure. The discussion is conducted in reference to the uncoupled case represented in Fig. 2a.

The uncoupled modes of Fig. 2a. are modeled by assimilating the TO Raman cross section with the imaginary part of the relative dielectric function $\varepsilon_r$ (that captures the $\varepsilon_r \to \infty$ divergence characteristic of a purely-mechanical TO)[64]. The adopted four-mode $\{2 \times (Cd-Te), 2 \times (Be-Te)\}$ description conforms with the current *ab initio* findings – referring to the end $i$-to-$iv$ modes. A sensitivity of both bond vibrations to their local (CdTe- and BeTe-like) environments at the first-neighbor scale is assumed by analogy with $Zn_{1-x}Be_xTe$[40], impacting the Raman intensities. The mechanical bond force constants are linearly interpolated between the parent and impurity *ab initio* values, leading to quadratic $x$-dependencies of the TO frequencies. A uniform broadening (1 cm$^{-1}$) is used, so that the Raman intensities (color code) can be directly compared. The remaining input parameters are determined *ab initio* (Sec. SIII.2); no adjustable parameters are used. The *ab initio* frequencies are slightly shifted upwards with respect to the experimental ones (symbols – Fig. 2a) – by a few cm$^{-1}$ for pure CdTe (hollow symbols) and by less than 15 cm$^{-1}$ for pure BeTe (filled symbols) – due to a generally known bias of the local density approximation, innate to our *ab initio* calculations, to overbind and hence to overestimate the bond force constants – except for the Be-Te impurity–$i$ mode at $x \sim 0$ (hollow symbols) for which a nearly perfect matching occurs.

Fair modeling of the pressure-induced interplay between the *ab initio* Raman intensities of the two Be-Te submodes forming the irregular-inverted $(TO_{Be-Te}^{Cd}, TO_{Be-Te}^{Be})$ percolation doublet of $Cd_{1-x}Be_xTe$ near the crossing of percolation branches ($x \sim 1$, Fig. 2e) can be achieved within a dielectric parametrization of two like $(TO_{Be-Te}^{-}, TO_{Be-Te}^{+})$ mechanically-coupled harmonic oscillators (Fig. 2f) – ranked in order of frequency, by slightly adapting the approach developed in Ref.[40] – notably using there a simplified form of Eq. (7) a – see Sec. SIII.2. The pressure dependence of the parent BeTe oscillator strength is taken into account as done in Ref.[40]. For better visualization of trends, the dielectric study has been moved from dilute ($x \sim 1$) to minor ($x=0.81$) Cd content – further offering a direct comparison with the regular $Zn_{1-x}Be_xTe$ case (referring to Fig. 1c of Ref.[40]). A small mechanical coupling is used (with characteristic frequency $\omega'=50$ cm$^{-1}$, similar to that used with ZnBeTe[40]) scaled down by (roughly) an order of magnitude with respect to the TO frequencies (~480 cm$^{-1}$) of the raw-uncoupled $(TO_{Be-Te}^{Cd}, TO_{Be-Te}^{Be})$ oscillators (whose pressure dependence are represented by straight and dotted lines in Fig. 2f, correspondingly). The small coupling successfully mimics the minimal – but finite – *ab initio* splitting manifesting the anticrossing (~3 cm$^{-1}$, emphasized by paired arrows in Fig. 2f) of the coupled $(TO_{Be-Te}^{-}, TO_{Be-Te}^{+})$ near the resonance – corresponding to frequency matching of the raw-uncoupled oscillators. At this limit, the mechanical coupling achieves maximum (~20 GPa) and the two Be-Te TO submodes exhibit comparable Raman intensities. A further pressure increase eventually results in a proper doublet inversion (30 GPa, Fig. 2f) – as the oscillators tend to decouple by shifting away from the resonance. Such inversion manifests, in fact, the free coupling of Be-Te oscillators at the resonance in $Cd_{1-x}Be_xTe$ – cf. the issue 4.

The free Be-Te coupling is also observed with $Zn_{1-x}Be_xTe$[40] ($x$~1), only that the transfer of Be-Te oscillator strength mediated by the mechanical coupling is opposite (due to inversion of the native Be-Te Raman doublets with respect to $Cd_{1-x}Be_xTe$ at 0 GPa – see above). A further difference relates to the Be-Te convergence rate under pressure at $x$~1, being fast for $Zn_{1-x}Be_xTe$ (the inversion is completed already at ~10 GPa, Ref.[40]) and slow for $Cd_{1-x}Be_xTe$ (the inversion is still in progress at the largest tested pressure of 20 GPa in our *ab initio* data – Fig. 2e, being delayed to ~30 GPa in view of our dielectric parametrization of Raman lineshapes – Fig. 2f). This reflects different driving forces behind the pressure-induced Be-Te convergence processes, namely the regular $\partial f_i/\partial\, lnV$-mechanism for $Zn_{1-x}Be_xTe$ – cf. the issue 3, and for $Cd_{1-x}Be_xTe$ just a basic trend for like bonds to behave uniformly under pressure (the local environment becoming less discriminatory at high pressure[40], as already mentioned). Anyway, the free coupling of the inverted / irregular Be-Te doublet of $Cd_{1-x}Be_xTe$ at large Be content (Figs. 2e and 2f) opposes the phonon exceptional point achieved at minor-to-moderate Be content (Figs. 2b, 2c and 2d), conforming to scenarii 2 and 1 in the main lines – cf. the issue 4.

## Conclusion

Among Be-based highly mismatched alloys (HMA's), $Cd_{1-x}Be_xTe$ achieves maximum contrast in bond properties, exacerbating its vibrational-mechanical properties. As such, $Cd_{1-x}Be_xTe$ is an appealing benchmark to test the limit of the percolation model (PM) that so far provided a unified understanding of the Raman spectra of semiconductor alloys – covering well matched alloys (WMA's) as well as HMA's.

The $Cd_{1-x}Be_xTe$ mechanical-vibrational properties are studied experimentally ($x \leq 0.11$) by addressing the bulk modulus $B_0$ and the effective bond force constants, using high-pressure X-ray diffraction and high-pressure Raman scattering in combination. The discussion of experimental data is supported by high-pressure *ab initio* (SIESTA, AIMPRO) snapshots at moderate Be contents ($x \leq 0.3$), extended to intermediary ($x$~0.5) and large ($x$~1) Be contents situated beyond the experimental stage.

$Cd_{1-x}Be_xTe$ exhibits degraded mechanical-vibrational properties across the composition domain, at both the macroscopic and microscopic scales. On minor Be incorporation ($x$~0), the bulk modulus suffers a counterintuitive drop below the CdTe value. At large Be content ($x \sim 1$) and ambient pressure, the percolation-type Raman signal of the sensitive Be-Te bond is fully relaxed by the phonon dispersion, down to inversion – an unprecedented case among revisited alloys within the PM so far. Yet, the percolation scheme is not challenged for all that and basically applies as such to $Cd_{1-x}Be_xTe$. In fact, the Be-Te bond exhibits a mere bimodal Raman pattern at any composition, and not a more complicated one. Further, the pressure-induced convergence of the Be-Te Raman doublet either ends up into a phonon exceptional point at minor-to-moderate Be content or develops into a free mechanical coupling at large Be content, in line with predictions. This reinforces the status of the PM as a robust (phenomenological) descriptor of phonons in semiconductor alloys.

**Methods**

This Section provides details on the samples and various experimental techniques and simulation (numerical and analytical) methods used to probe and / or to support the discussion on the structural, electronic and / or vibrational properties of $Cd_{1-x}Be_xTe$, depending on pressure.

**Samples.** $Cd_{1-x}Be_xTe$ ($x$=0, 0.03, 0.05, 0.07, 0.11) and $Zn_{1-x}Be_xTe$ ($x$=0.04, 0.21) free-standing single crystals required for high-pressure (Raman scattering and / or X-ray diffraction) measurements are grown from the melt by mixing high-purity CdTe and ZnTe (99.9995, *i.e.*, 5N quality) with Be (99.5, *i.e.*, 2N quality) using the Bridgman method[65]. The $Cd_{1-x}Be_xTe$ composition is determined better than 1% by selective Cd and Be dosing via the inductively coupled plasma (ICP) method applied to powders. All samples crystallize in the native zincblende structure of CdTe and BeTe, verified by powder X-ray diffraction measurements at 0 GPa – Sec. S.I). The $Zn_{1-x}Be_xTe$ composition is derived from the linear composition dependence of the lattice constant[46] measured at 0 GPa by powder X-ray diffraction.

**Solid-state nuclear magnetic resonance measurements.** Solid-state nuclear magnetic resonance measurements of the chemical shift due to the invariant Te species of $Cd_{1-x}Be_xTe$ are performed on a finely ground powder (~20 ml) obtained from the largest crystal piece at hand ($x$=0.07) to test whether the Cd↔Be atom substitution is ideally random, or not (*i.e.*, subject to clustering or anticlustering). For all solid-state NMR spectra, Bruker 2.5 mm double resonance probes are used and the magic angle spinning frequency is set to 25 kHz. $^{125}Te$, $^{113}Cd$ and $^{9}Be$ longitudinal relaxation times ($T_1$), respectively 450 s, 840 s and 175 s, are measured using standard saturation recovery experiments. Recycle delays of 5 times $T_1$ are then used to ensure quantitativity. $^{125}Te$ NMR spectrum is acquired on a Bruker Avance III spectrometer operating at 7 T (corresponding to 300 MHz $^{1}H$ resonance frequency). The direct-acquisition Carr-Purcell-Meiboom-Gil (CPMG) experiment[66] enables to record, in each scan, 49 full echoes with 400 ms acquisition time. 256 scans are accumulated in 7 days and 9 hours. $^{113}Cd$ and $^{9}Be$ NMR standard direct acquisition experiments are recorded on a Bruker Avance III 600 MHz spectrometer (14 T). Respectively 112 and 56 transients are recorded, for a total experimental time of 5 days 10 h for $^{113}Cd$ and 15 h for $^{9}Be$.

**Ellipsometry measurements.** Near the direct absorption edge, the square root of the deviation ($E - E_0$) from the fundamental band gap ($E_0$) scales linearly with the energy ($E$) weighted by the absorption coefficient ($\alpha$). $E_0$ is obtained by measuring such so-called Tauc plots[67] in the visible using non-oriented crystal pieces with parallel faces polished to optical quality – both directly, in a transmission experiment at normal incidence, and indirectly, by ellipsometry at a near-Brewster incidence (using a HORIBA UVISEL phase modulated spectroscopic ellipsometer). Higher interband transitions ($E_1, E_1 +$ $\Delta_1$, $E_2$) on top of $E_0$ are accessed via a direct (model-free) wavelength-per-wavelength inversion of spectrometric ellipsometry data (the sine and cosine of the depolarization angles).

**(High-pressure) X-ray diffraction measurements.** High-pressure powder X-ray diffractograms are recorded on the PSICHÉ ($Cd_{0.89}Be_{0.11}Te$) and CRISTAL ($Zn_{1-x}Be_xTe$, $x$=0.045, 0.14 and 0.21) beamlines of SOLEIL synchrotron using radiation wavelengths of 0.3738 Å and 0.485 Å, respectively. The high-pressure data are recorded with a similar Chervin type diamond anvil cell[68] (with 300 µm in diameter diamond culet) as for the Raman measurements (see below), using, in both cases, Ne and Au as the pressure transmitting medium and for pressure calibration, respectively. The high-pressure X-ray data are treated by using the DIOPTAS[69] software at PSICHÉ and the DATLAB software – kindly provided by K. Syassen (Max-Planck Institut für Festkörperphysik, Stuttgart, Germany) – at CRISTAL. The $B_0$ value of the native zincblende phase at 0 GPa is obtained by fitting the Birch-Murnaghan's equation of state to the pressure dependence of the unit cell volume[51]. The volume of the unit cell at 0 GPa coming into the cited equation is determined from powder X-ray diffraction measurement done at 0 GPa in laboratory conditions using the CuK$\alpha$ radiation.

**(High-pressure) Raman scattering measurements.** High-pressure / low-temperature Raman experiment on $Cd_{0.89}Be_{0.11}Te$ is done in the backscattering geometry by focusing the laser beam through a ×50 microscope objective with a long working distance at normal incidence onto a 40 µm in diameter spot at the (non-oriented) surface (polished to optical quality) of a tiny monocrystal inserted into a Chervin type diamond anvil cell (described above) set into a Helium flow cryostat system operated at liquid nitrogen temperature. CdTe-based crystals being notoriously poor Raman scatterers[59], near-resonant conditions are used to enhance the Raman signal, as best achieved at 300 K and 77 K by using the 632.8 nm line of a helium-neon laser, falling in between $E_0$ and $E_0 + \Delta_0$, and the 488.0 nm line of an argon laser, near-resonant with $E_0 + \Delta_0$, respectively. Fair modeling of the Raman signal due to uncoupled or mechanically-coupled TO oscillators is achieved within a linear dielectric function approach, adapted from Ref.[40] (Sec. SIII).

**(High-pressure) *ab initio* insights into the lattice relaxation/dynamics.** Three *ab initio* codes operated within the density functional theory, using pseudopotentials and the local density approximation for the exchange-correlation, are employed, according to need. Reference $B_0$ vs. $x$ curves for $Cd_{1-x}Be_xTe$ at moderate Be content ($x \leq 0.3$) are separately obtained by applying the SIESTA[43,44] and AIMPRO (Ab Initio Modeling PROgram[41,42]) codes to distinct series of large (64- and 216-atom, respectively) fully-relaxed (lattice constant, atom positions, supercell shape) and partially-relaxed (preventing any supercell distortion so as to maintain the cubic structure needed for a strict application of the Birch-Murnaghan[51] equation of state) disordered zincblende-type supercells, correspondingly. Each supercell represents a random Cd↔Be substitution, obtained by adjusting the fractions of individual Te-centred tetrahedra to the binomial Bernouilli distribution[39]. At each $x$ value, $B_0$ is estimated by fitting the pressure dependence of the volume unit cell with the Birch-Murnaghan equation of state[51]. Further AIMPRO calculations are done on parent-like supercells containing Be-($x$~0) or Cd-paired ($x$~1) impurity motifs, searching to imitate the limiting-cases Raman signals due to the non-polar (purely-mechanical) TO's of the host and impurity species, using the formula given by de Gironcoli[70]. In their current versions, the AIMPRO and SIESTA codes do not take into account the long-range electric field accompanying the Raman-active polar LO's. Further simulations thus resorted to QE code[45], giving access to $\omega_T$ (the frequency of the non-polar TO), $\omega_L$ (the frequency of the Raman-active polar LO) and $\varepsilon_\infty$ (the high-frequency relative dielectric constant, *i.e.*, as defined at $\omega \gg \omega_T$) of BeTe and CdTe, calculated for (2×2×2) cubic zincblende-type supercells (64 atoms). From these, the static relative dielectric constant $\varepsilon_s$ (defined at $\omega \ll \omega_T$) could have been extracted by force of the LST relation[71], to be further used to estimate the phonon oscillator strengths within the classical form of the relative $Cd_{1-x}Be_xTe$ dielectric function $\varepsilon_r$, used in our simplified analytical expression of the $Cd_{1-x}Be_xTe$ Raman cross section (Sec. SIII).

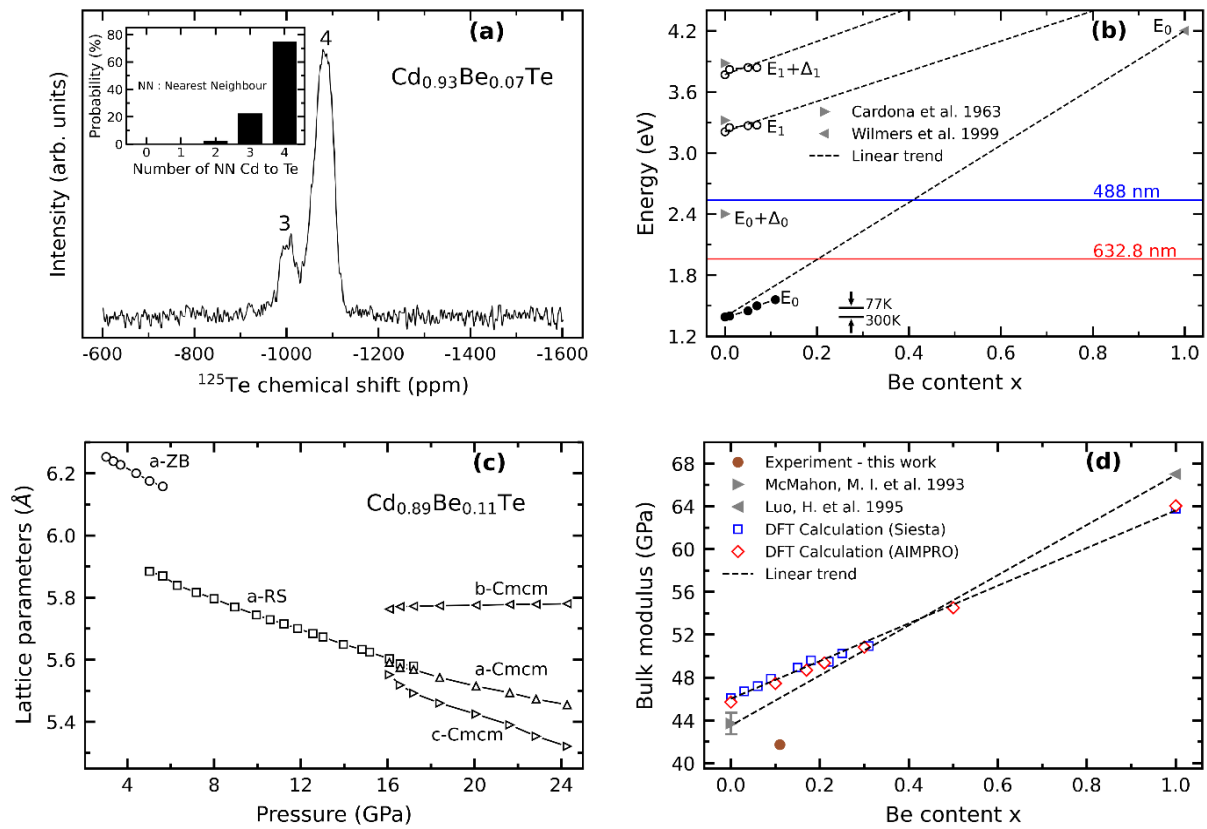

**Figure 1 | $Cd_{1-x}Be_xTe$ structural, optical and mechanical properties.** (a) CPMG $Cd_{0.93}Be_{0.07}Te$ $^{125}Te$ NMR signal. The binomial distribution of Te-centered nearest-neighbor (NN) tetrahedon clusters depending on the number of Cd atoms at the vertices in case of a random Cd↔Be substitution is added for comparison (inset). The NMR peaks are labeled accordingly. (b) Composition dependence of the main $Cd_{1-x}Be_xTe$ electronic transitions measured at room temperature by transmission (filled symbols, Fig. S5a) and ellipsometry (hollow symbols, Fig. S5b). CdTe (Ref.[48]) and BeTe (Ref.[20]) values taken from the literature are added, for reference purpose. Linear (dashed) trends between parent values are guidelines for the eye. Laser lines used to excite the Raman spectra are positioned to appreciate resonance conditions. Antagonist arrows help to appreciate the shift of electronic transitions by lowering temperature from ambient to liquid nitrogen, by referring to the $E_0$ gap of CdTe Ref.[49]. (c) Pressure dependence of the zincblende (zb), rocksalt (rs) and Cmcm (cm) $Cd_{0.89}Be_{0.11}Te$ lattice constant(s) measured by high-pressure X-ray diffraction (Fig. S1c). (d) The $B_0$ value derived for $Cd_{0.89}Be_{0.11}Te$ in its native zb phase (filled circle) from the corresponding volume vs. pressure dependence (Fig. S1d) is compared with the parent values taken from the literature (filled triangles, Refs.[30,50]) and with current *ab initio* data obtained with the AIMPRO (hollow diamonds) and SIESTA (hollow squares) codes. Corresponding linear $x$-dependencies are shown (dashed lines), for reference purpose.

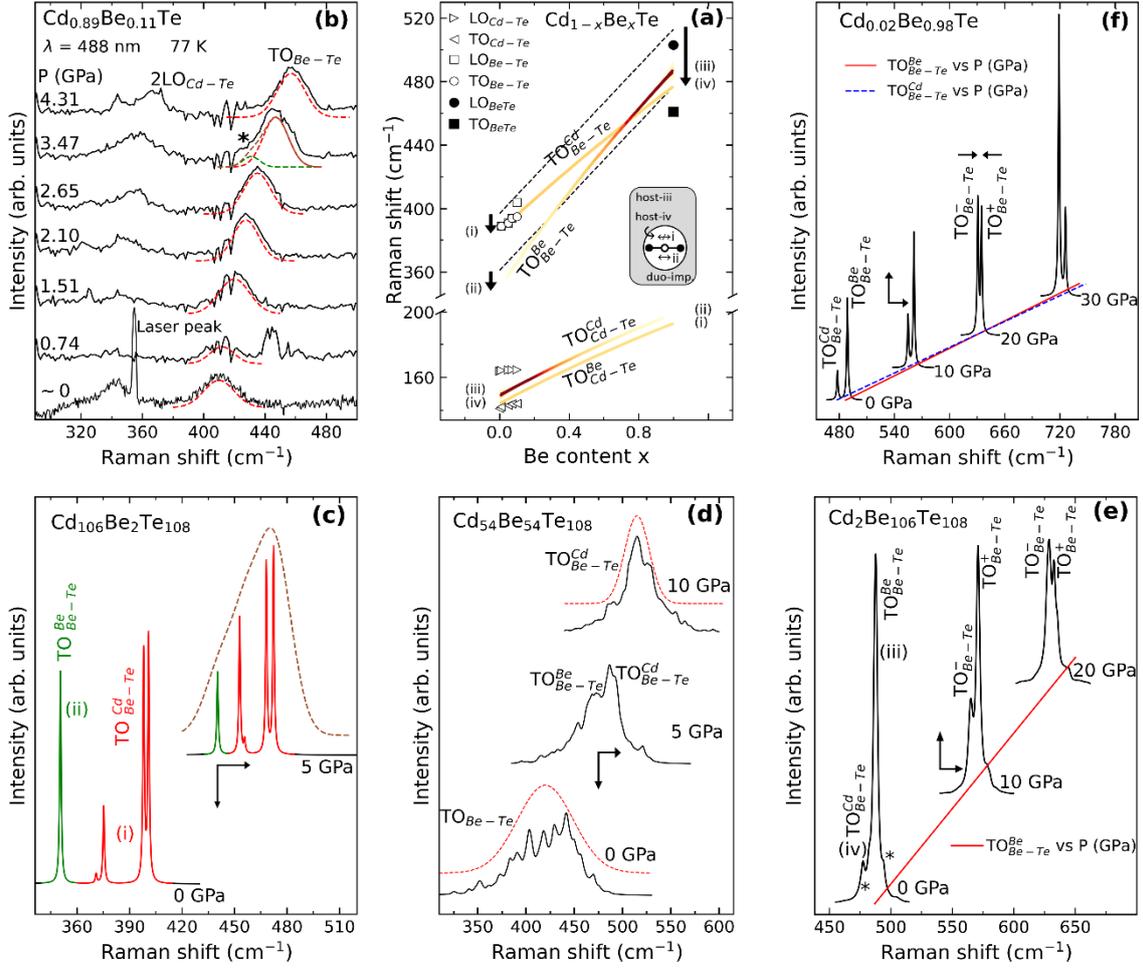

**Figure 2 | Cd$_{1-x}$Be$_x$Te vibrational properties.** Panels are arranged anti-clockwise – starting from an overview at the top-center **(a)** – in the sense of increasing $x$ values from left ($x\sim0$) to right ($x\sim1$) for direct vertical comparisons of the side panels, i.e., **(b)** vs. **(c)** and **(f)** vs. **(e)**, with a snapshot at intermediary composition in-between **(d)**. **(a)** Theoretical overview of the Cd$_{1-x}$Be$_x$Te TO Raman frequencies (curves) and intensities (color of curves) at 0 GPa within a four-mode $\{2 \times (Cd-Te), 2 \times (Be-Te)\}$ description in absence of mechanical coupling between oscillators ($\omega'=0$). A sensitivity of bond vibrations to first neighbors is assumed by analogy with Zn$_{1-x}$Be$_x$Te[40]. The current TO and LO Raman frequencies (hollow symbols) are indicated. BeTe data taken from the literature[55] are added (filled symbols), for reference purpose. A schematic view of the prototypical parent-like supercell (216-atom) containing one isolated impurity-duo used to generate an ab initio insight into the end ($x\sim0,1$) TO Raman frequencies of Cd$_{1-x}$Be$_x$Te is inserted. Labels (i) to (iv) refer to in-chain, out-of-chain, near-chain and away-from-chain bond vibrations – in this order, as sketched out. The shape of the Be-Te doublet dictated by the sole effect of the local strain, i.e., in absence of dispersion, is schematically represented by straight-dashed parallel lines. The dispersion effect affecting the impurity modes is emphasized by vertical arrows at $x\sim0,1$. **(b)** High-pressure / low-temperature Cd$_{0.89}$Be$_{0.11}$Te Raman spectra taken in the upstroke. The Be-Te signal, modeled via Lorentzian functions (dotted lines) transiently exhibits a minor feature at ~3.5 GPa (marked by an asterisk). **(c, d, e)** Ab initio (AIMPRO) insights into the pressure dependence of the Be-Te Raman signal in large (216-atom) supercells **(c)** due to the Be-duo ($x\sim0$), **(d)** at intermediary Be content ($x=0.5$) and **(e)** in presence of the Cd-duo ($x\sim1$) – giving rise to various local modes (spotted by asterisks). **(f)** Raman cross section reflecting the pressure-dependence of the irregular-inverted (see text) Be-Te TO Raman doublet at minor Cd content ($x\sim0.81$) in case of a minor mechanical coupling ($\omega'=50$ cm$^{-1}$). Straight and dotted lines represent the pressure-dependencies of the frequencies of the raw-uncoupled Be-Te modes behind the coupled ones. In panels **(b, c)**, a color code is used to distinguish between the raw-uncoupled TO's stemming from domestic (red) and foreign (blue) environments. Paired vertical – horizontal arrows (in panels **(b – f)** emphasize pressure-induced changes in Raman intensity – frequency for a given mode.

**Data availability statement**

The reported experimental (high-pressure Raman, high-pressure X-ray diffraction, nuclear magnetic resonance) and *ab initio* (AIMPRO, SIESTA and QE codes) data in this work are available upon request to the corresponding author.

**Acknowledgements**

We acknowledge assistance from the PSICHÉ ($Cd_{1-x}Be_xTe$, Proposal n°. 20220436, co-resps. O.P. & A.P.) and CRISTAL ($Zn_{1-x}Be_xTe$ Proposal n°. 20110018, resp. A.P.) beamlines of synchrotron SOLEIL for the high-pressure X-ray diffraction measurements, from the Plate-forme de Spectroscopie – Institut de Minéralogie, de Physique des Matériaux et de Cosmochimie – Sorbonne Université (http://impmc.sorbonne-universite.fr/fr/plateformes-et-equipements/plate forme de spectroscopie.html), from the Nuclear Magnetic Resonance platform of Institut Jean Barriol (CNRS FR 3843, Université de Lorraine – https://www.ijb.univ-lorraine.fr), from the Ellipsometry core facility of LCP-A2MC (Université de Lorraine – https://lcp-a2mc.univ-lorraine.fr) and from the IJL core facility (Université de Lorraine – http://ijl.univ-lorraine.fr/recherche/centres-de-competences/rayons-X-et-spectroscopie-moessbauer-X-gamma) for the X-ray diffraction measurements at ambient pressure. The SIESTA and QE calculations have been done using the facilities of the EXPLOR mesocentre of the Université de Lorraine (project 2019CPMXX0918). VJBT acknowledge the FCT through projects LA/P/0037/2020, UIDB/50025/2020 and UIDP/50025/2020 in relation to the AIMPRO calculations.

**Affiliations**

Université de Lorraine, LCP-A2MC, ER 4632, F-57000 Metz, France

A. Elmahjoubi, M. B. Shoker, O. Pagès, A.V. Postnikov, A. EnNaciri, L. Broch, R. Hajj Hussein & P. Franchetti

Departamento de Fisica and i3N, Universidade de Aveiro, 3810 – 193 Aveiro, Portugal

V. J. B. Torres

Institut de Minéralogie, de Physique des Matériaux et de Cosmochimie, Sorbonne Université — UMR CNRS 7590, F-75005 Paris, France

A. Polian, C. Bellin & K. Beneut

Université de Lorraine, Laboratoire de Cristallographie, Résonance Magnétique et Modélisations, CRM2, UMR 7036, Vandoeuvre-lès-Nancy, F-54506, France

C. Gardiennet & G. Kervern

Synchrotron SOLEIL, L'Orme des Merisiers Saint-Aubin, BP 48 F-91192 Gif-sur-Yvette Cedex, France

J.-P. Itié, L. Nataf & S. Ravy

Université de Lorraine, CNRS, IJL, F-57000, Metz, France

S. Diliberto & S. Michel

Faculty of Physics, Astronomy and Informatics, Institute of Physics, Nicolaus Copernicus University in Toruń, ul. Grudziądzka 5, 87-100 Toruń, Poland

A. Abouais & K. Strzałkowski

Chouaib Doukkali University of El Jadida, National School of Applied Sciences, Engineering Science for Energy Lab, El Jadida, Morocco

A. Abouais

**Author Contributions Statement**

High-quality / purity $Cd_{1-x}Be_xTe$ ($x \leq 0.11$) and $Zn_{1-x}Be_xTe$ ($x \leq 0.21$) free-standing single crystals with zincblende structure were grown by using the Bridgman method (A.A. & K.S.). The Cd↔Be substitution was checked to be ideally random by applying solid-state nuclear magnetic resonance to a (large) powdered sample ($x$=0.07 – G.K. & C.G.). The composition of $Cd_{1-x}Be_xTe$ alloys was determined by using the induced coupling plasma method (S.M.). X-ray diffraction measurements on powders were done in laboratory at ambient pressure across the sample series (S.D.) – to check the crystal structure and to determine the $x$-dependence of lattice constant(s) – and under pressure at the PSICHÉ ($Cd_{1-x}Be_xTe$ – A.P., M. B. S. & O. P. with the assistance of J.-P.I) and CRISTAL ($Zn_{1-x}Be_xTe$ – A.P., R. H. H. & O. P. with the assistance of L.N. & S.R.) beamlines of SOLEIL synchrotron – to determine the pressure domain of the native zincblende phase together with sequence of pressure-induced structural transitions. The high-pressure X-ray data were subsequently analyzed using the software DIOPTAS (A.E.) and DATLAB (M.B.S. & R.H.H.). The Be-Te Raman signal of $Cd_{1-x}Be_xTe$ was studied at ambient-pressure / low-temperature (A.E., M.B.S & O.P., with the assistance of P. F.) and high-pressure / low-temperature ($x$=0.11 – A.P., C.B., K.B., A. E. & O.P.) using single crystals. *Ab initio* calculations of high-pressure Raman spectra, bulk modulus ($B_0$) and various parameters ($\varepsilon_\infty$, TO-LO splitting) needed to implement the linear dielectric function approach were done by using AIMPRO (V.J.B.T.), SIESTA (A.E. & A.V.P.) and QE (A.E.). Phenomenological modeling of the Raman cross section for a coupled system of harmonic oscillators within the linear dielectric function approach (O.P.) was implemented on MATLAB (A.E. & M.B.S.). O.P. wrote the manuscript, with the help of A.V.P and contributions from all co-authors. The figures were done by A.E.

**Additional Information**

Supplementary information accompanies this paper at https://...

**Competing interests**

The authors declare no competing interests.

**Corresponding author**

Correspondence to Olivier Pagès.

**Supplementary Information**

**Vibrational-mechanical properties of the highly-mismatched Cd$_{1-x}$Be$_x$Te semiconductor alloy – Experiment and *ab initio* calculations**

A. Elmahjoubi, M. B. Shoker, O. Pagès, V. J.B. Torres, A. Polian, A.V. Postnikov, C. Bellin, K. Beneut, C. Gardiennet, G. Kervern, A. EnNaciri, L. Broch, R. Hajj Hussein, J.-P. Itié, L. Nataf, S. Ravy, P. Franchetti, S. Diliberto, S. Michel, A. Abouais and K. Strzałkowski

In this annex, supplementary experimental, *ab initio* (AIMPRO) and modeling (linear dielectric approach) material is presented further supporting the discussion of the structural, optical (**Sec. SI**) and mechanical-vibrational properties of Cd$_{1-x}$Be$_x$Te (**Sec. SII**) done in the main text. Raw experimental data concerned with the structural properties of Be-dilute Cd$_{1-x}$Be$_x$Te studied at the lattice-macroscopic and bond-microscopic scales by high-pressure X-ray diffraction (HP-XRD) and by solid state nuclear magnetic resonance (SS-NMR) – completing Fig. 1 – are reported in **Secs. SI.1** and **SI.2**, correspondingly, including a fruitful comparison with Zn$_{1-x}$Be$_x$Te in the former case (HP-XRD, **Sec. SI.1**). **Sec. SI.3** reports on the main Cd$_{1-x}$Be$_x$Te electronic interband transitions ($E_0, E_1, E_1 + \Delta_1$) determined by combining transmission and ellipsometry measurements in the visible. **Sec. SII.** provides *ab initio* insights into the Cd$_{1-x}$Be$_x$Te lattice relaxation for an isolated impurity (**Sec. II.1**), for paired impurities (**Sec. II.2**) and at intermediate composition (**Sec. II.3**). In **Sec. II.1**, an isolated Be impurity in CdTe – suffering a hydrostatic tensile strain from the host matrix – offers a benchmark to estimate in a full *ab initio* (SIESTA) approach how the local strain and the phonon dispersion separately impact the Raman frequencies of impurity modes. **Sec. II.2** completes the high-pressure *ab initio* Raman study of paired-impurities in parent-like supercell done in the main text (Figs. 2c and 2e) by shifting the focus to the Cd-Te spectral range. **Sec. II.3** provides an *ab initio* insight into the pressure dependence of the lattice relaxation within the Cd$_{54}$Be$_{54}$Te$_{108}$ cubic supercell used to calculate the *ab initio* high-pressure Raman spectra (Fig. 2d). **Sec. SIII** is concerned with the Cd$_{1-x}$Be$_x$Te lattice dynamics. **Sec. III.1** reports on an experimental Raman study of the Cd$_{1-x}$Be$_x$Te vibrational properties at minor Be contents done at 0 GPa – in preparation for the high-pressure Raman study (Fig. 2b). Last, in **Sec. SIII.2** we outline the linear dielectric approach used for contour modeling of the various Raman lineshapes (Fig. 2), referring to the four-mode $\{TO_{Cd-Te}^{Be}, TO_{Cd-Te}^{Cd}, TO_{Be-Te}^{Be}, TO_{Be-Te}^{Cd}\}$ description across the composition domain at 0 GPa in absence of mechanical coupling ($\omega'$=0 cm$^{-1}$, Fig. 2a), and to the punctual insight into the pressure dependence of the Be-Te Raman doublet at $x\sim1$ achieved by considering a weak mechanical coupling ($\omega'$=50 cm$^{-1}$, Fig. 2f) impacting both the Raman frequencies and intensities.

I. **Structural and optical properties of Be-dilute Cd$_{1-x}$Be$_x$Te**

I.1. (High-pressure) X-ray diffraction – Structural insight at the macroscopic (lattice) scale

Raw X-Ray diffractograms obtained at 0 GPa in laboratory using the CuK$\alpha$ radation across the current Cd$_{1-x}$Be$_x$Te ($x \leq 0.11$) sample series (Fig. S1a) are characterized by sharp peaks at any composition $x$ value, the sign of a high structural quality. The individual peaks are labelled using the (hkl) Miller indices up to maximal angular deviation. The lattice constant $a$ is found to vary linearly with the composition $x$ (Fig. S1b), a standard feature of semiconductor alloys[2].

A selection of raw high-pressure X-ray diffractograms taken at increasing pressure on Cd$_{0.89}$Be$_{0.11}$Te at the PSICHÉ beamline of SOLEIL synchrotron using the 0.3738 Å radiation (Fig. S1c) reveal no extra peak besides the regular ones for a given crystal phase, the sign of a high structural purity. As pressure increases, Cd$_{0.89}$Be$_{0.11}$Te transforms from zincblende (0 GPa, abbreviated zb) to rocksalt (~5.5 GPa, rs) – the two phases coexisting over a rather large pressure domain (~5.5 – 9 GPa) – and subsequently to Cmcm (~13 GPa, cm), eventually surviving as a unique phase from ~18 GPa up to the maximum achieved pressure in this study (~24 GPa). The pressure dependence of lattice constants in each Cd$_{0.89}$Be$_{0.11}$Te phase (Fig. 1c) is notably used to estimate the bulk modulus $B_0$ at 0 GPa in the native zincblende phase (Fig. 1d) by fitting the pressure dependence of the volume of the unit cell to the Birch-Murnaghan equation of state[51] (Fig. S1d).

Fig. S2 displays selections of similar high-pressure X-ray diffractograms obtained with Zn$_{1-x}$Be$_x$Te at (a) $x$=0.045 and (b) 0.21 at the CRISTAL beamline of SOLEIL synchrotron using the 0.485 Å radiation – completing the already published $x$=0.14 data[40] taken during the same run of experiment. The related experimental (symbols) volume versus pressure dependencies in the native zincblende phase fitted to the Birch-Murnaghan equation of state[51] (curves) and resulting $B_0$ vs. $x$ variations are displayed in Figs. S3a and S3b, respectively. The fit is done by fixing the reference volume at 0 GPa to the value obtained by considering a linear dependence of the lattice constant versus $x$ in Zn$_{1-x}$Be$_x$Te[46]. The as-fitted $B_0$ (symbols) and $B_0'$ values at $x$=(0.045, 0.14, 0.21) are (52.857±0.640, 52.846±0.170, 52.542±0.150) in GPa (the error bars is within the symbol size) and (3.94±0.18, 4.05±0.09, 4.03±0.06), respectively. For all mixed crystals $B_0'$ remains stable at around 4, found relevant for ZnTe[54] and BeTe[53]. A pronounced deviation from the $B_0$ vs. $x$ linearity (dashed line in Fig. S3b) in the sense of a negative bowing is observed – as in the case of Cd$_{1-x}$Be$_x$Te (Fig. 1d) – with a minimal $B_0$ value at 21 at.% Be, quasi matched with the parent ZnTe value[54] – 52 GPa, well below the BeTe one[53] – 67 GPa.

### I.2. Solid state nuclear magnetic resonance (SS-NMR) – Structural insight at the microscopic (atom) scale

In an alloy with zincblende structure such as $Cd_{1-x}Be_xTe$, the substituent (Cd and Be in this case) and invariant (Te) species are intercalated so a to form a (cubic) tetrahedral arrangement. Hence, Cd and Be exhibit a stable nearest-neighbor environment of four Te atoms at any $x$ value. In contrast the tetrahedral environment of Te diversifies into five variants depending on the number of Be and Cd atoms at the vertices. The five types of Te-centrered tetrahedra are present in the crystal at any $x$ value, with various probabilities depending on $x$, following the binomial Bernoulli's distribution in their $x$-dependence in the ideal case of a random Be↔Cd substitution[S1].

The pioneering $^{125}$Te solid-state nuclear magnetic resonance (NMR) measurements performed on $Cd_{1-x}Zn_xTe$ by Zamir *et al.*[39] have demonstrated a sensitivity of the NMR shift to the local environment at the nearest-neighbor scale. Hence, only the NMR data related to the invariant Te species (as opposed to the substitutional Cd and Be ones) of $Cd_{1-x}Be_xTe$ can shed light on the nature of the Cd↔Be atom substitution – the reason for the emphasis on Te in the main text (Fig. 1a). However, the $^9$Be (a) and $^{113}$Cd (b) NMR data related to both substituents are also provided (Fig. S4), for the sake of completeness. A unique (well-defined) feature, reflecting the uniqueness of the local environment (all-Te), is visible in each case, as ideally expected.

### I.3. Optical properties – Transmission and ellipsometry in combination

The main $Cd_{1-x}Be_xTe$ electronic interband transitions are determined across the sample series in their $x$-dependence by combining transmission ($E_0$) and ellipsometry ($E_1, E_1 + \Delta_1, E_2$) measurements in the visible (Fig. 1b). Selected data (Fig. S5) illustrate how the interband transitions are estimated in practice from the raw spectrometric data at each composition, *i.e.*, via a Tauc plot (transmission data, Fig. S5a – curve, filled symbols in Fig. 1b) or by direct (model free) inversion of the measured depolarization angles giving access to the imaginary part of the dielectric function within the 0.6 – 5.5 eV spectral range (ellipsometry data, Fig. S5b, hollow symbols in Fig. 1b).

## II. $Cd_{1-x}Be_xTe$ lattice relaxation / dynamics – *Ab initio* insights

### II.1. $Cd_{31}Be_1Te_{32}$ (SIESTA code) – isolated-impurity motif

An isolated Be atom in CdTe forms short Be-Te bonds suffering a hydrostatic tensile strain from CdTe corresponding to a large bond length. The Be-Te bond elongation $\Delta l$ with respect to the natural bond length in the pure BeTe crystal ($l_0$) softens the Be-Te impurity mode in CdTe – referred to as the

impurity-$i$ mode in the main text – below the parent BeTe TO frequency ($\omega_{T,0}$). The shift in TO frequency squared $\Delta\omega_T^2$ relates to the variation in bond length $\Delta l$ via the relation[63,S2],

$$\frac{\Delta\omega_T^2}{\omega_{T,0}^2} = -6\gamma_T \cdot \frac{\Delta l}{l_0}, \qquad (1)$$

that involves the Grüneisen parameter of the parent TO mode, given by[S3],

$$\gamma_T = \frac{B_0}{\omega_{T,0}} \cdot \left(\frac{d\omega_T}{dP}\right)_{P=0}, \qquad (2)$$

where $P$ is the hydrostatic pressure and $B_0$ the bulk modulus.

High-pressure *ab initio* (SIESTA) calculations of the lattice relaxation and of the Γ-projected phonon density of states – that assimilates with the TO Raman spectrum in a crude approximation – done on a large 2×2×2 (64-atom) fully relaxed ($l_0$~2.406 Å at ambient pressure) cubic BeTe supercell yield $B_0$ and $\omega_{T,0}$ values of 63.7 GPa and 478 cm$^{-1}$ (nearly matching the AIMPRO values in Figs. 1d and 2a), respectively, and further predict a linear increase of the TO frequency with pressure at low values (in the range 0 – 20 GPa) at the rate of ~10.05 cm$^{-1}$ *per* GPa. The resulting $\gamma_T$ estimate via Eq. (2) for the parent BeTe TO mode is ~1.33.

Similar SIESTA calculations done at 0 GPa on a similar zincblende-type Cd$_{31}$Be$_1$Te$_{32}$ supercell containing a unique Be atom give the Be-Te impurity-$i$ mode at ~414 cm$^{-1}$ (slightly above the experimental $TO_{Be-Te}^{Cd}$ one at $x$~0 – see Fig. 2a) after relaxation (achieved via a Be-Te elongation of $\Delta l$~0.063 Å with respect to the natural-bulk $l_0$ value – see above). This exceeds by ~10 cm$^{-1}$ the impurity-$i$ frequency due to the sole effect of the local strain given by Eq. (1). The difference is due to the (negative) dispersion of the BeTe TO mode, impacting all impurity ($i$-to-$iii$) modes (as emphasized by vertical arrows in Fig. 2a), in principle[60]. Such study of the lattice relaxation / dynamics around an isolated Be impurity in CdTe ($x$~0), consistently conducted end-to-end within an *ab initio* (SIESTA) approach in reference to BeTe ($x$~1), points towards the important role of the phonon dispersion besides the local strain in shaping the Be-Te percolation doublet of Cd$_{1-x}$Be$_x$Te.

II.2. Cd$_{106}$Be$_2$Te$_{108}$ and Cd$_2$Be$_{108}$Te$_{108}$ (AIMPRO code) – Cd-Te signal vs. paired-impurity motifs

Fig. S6 displays high-pressure *ab initio* (AIMPRO) CdTe-like TO Raman spectra due to large (216-atom) fully-relaxed (see methods) cubic Cd$_{106}$Be$_2$Te$_{108}$ (at 0 and 5 GPa – Fig. S6a) and Cd$_2$Be$_{108}$Te$_{108}$ (at 0 and 20 GPa – Fig. S6b) supercells containing similar (Be- and Cd) impurity-duos (connected via Te), in support to the discussion of the pressure dependence of the Cd-Te doublet of Cd$_{1-x}$Be$_x$Te (main text).

The Cd-Te doublet exhibits nearly the same spacing between the relevant ($iii$-$iv$, $x$~0) and ($i$-$ii$, $x$~1) Cd-Te modes at both ends of the composition domain. Under pressure the doublets either cross ($Cd_{106}Be_2Te_{108}$, Fig. S6a) or freeze into a phonon exceptional point at the resonance ($Cd_2Be_{106}Te_{108}$), already achieved at 0 GPa in this case – the frozen submode identified by its wavevector is spotted by an arrow (Fig. S6b).

### II.3. $Cd_{54}Be_{54}Te_{108}$ (AIMPRO code) – intermediary composition

Fig. S7 reports on *ab initio* (AIMPRO) calculations of the bond angle distribution depending on pressure within the large disordered zincblende-type $Cd_{54}Be_{54}Te_{108}$ ($x$=0.5) supercell used to generate the corresponding *ab initio* (AIMPRO) Raman spectra (Fig. 2d). The data reveal a prohibitive supercell distortion from 15 GPa onwards. This is manifested by a deviation of bond angles from the nominal zincblende value (109°), to such extent that two distinct groups of bond angles are identified (dotted lines). The zinclende structure is only preserved up to 10 GPa – being increasingly damped with pressure – thus fixing the limit for the $Cd_{54}Be_{54}Te_{108}$ *ab initio* Raman study (Fig. 2d). A similar collapse is suffered by the CdTe-like $Cd_{106}Be_2Te_{216}$ supercell containing the isolated Be-duo from 10 GPa onwards (not shown) used to provide an *ab initio* (AIMPRO) insight into the pressure dependence of the $Cd_{1-x}Be_xTe$ Raman spectra in the Be-dilute limit ($x$~0, Fig. 2c).

### III.     $Cd_{1-x}Be_xTe$ lattice dynamics – Raman scattering

### III.1.     Experimental Raman study at ambient pressure ($x \leq 0.11$)

Fig. S8 displays a selection of unpolarized $Cd_{1-x}Be_xTe$ ($x \leq 0.11$) Raman spectra taken at ambient pressure / temperature in the backscattering geometry on non-oriented crystal faces using the red (632.8 nm, $x$=0 and 0.5) and blue (488.0 nm, $x$=0.11) laser lines near-resonant with $E_0$ and $E_0 + \Delta_0$, respectively (Fig. 1b). The Raman signal consists of a quasi-degenerated TO-LO Be-Te impurity mode situated at much higher frequency than the CdTe-lattice TO-LO band (140 – 170 cm$^{-1}$). Near-resonance conditions favor the (polar) LO's compared with the (non-polar) TO's via the Fröhlich mechanism[S4,S5]. The LO emphasis is especially pronounced with the red laser excitation, manifested by the strong emergence of the second-order matrix signal ($2\times LO_{Cd-Te}$). A pure Be-Te TO-insight is searched for at the largest available Be content ($x$=0.11) by exciting a (110)-cleaved crystal face at normal incidence with the less resonant blue laser line, and collecting the scattered light in backscattering in the (TO-allowed, LO-forbidden) scattering geometry[S5].

The Raman spectrum at minimal Be content (5 at.%) provides a crude experimental estimate of the Be-Te impurity-*i* frequency in CdTe, *i.e.*, ~390 cm$^{-1}$, in close agreement with existing data in the literature[56,57]. Besides, the comparison between the LO-like (non-oriented crystal face) and TO-like (cleaved face) $Cd_{0.89}Be_{0.11}Te$ Raman spectra resolves a finite Be-Te TO-LO splitting (emphasized by dotted lines, Fig. S7). By decreasing $x$ the lattice constant $a$ increases (Fig. S1b) which distances cations and anions and hence softens the chemical bonds, with concomitant impact on the TO (non-polar) Raman frequencies (scaling as the square roots of the effective bond force constant in a crude description[26]) being downward shifted. The LO (polar) Raman frequencies "mechanically follow", converging progressively towards the (non-polar) TO's when the bond fractions decrease until the TO-LO degeneracy is achieved in the dilute limits ($x$~0 for Be-Te and $x$~1 for Cd-Te). An overview of the experimental $Cd_{1-x}Be_xTe$ Raman frequencies is given in Fig. 2a (symbols).

III.2. Contour modeling of (high-pressure) TO Raman spectra – linear dielectric function approach

An overview of the TO $Cd_{1-x}Be_xTe$ Raman frequencies (curves) and intensities (thickness of curves) calculated throughout the composition domain at 0 GPa in absence of mechanical coupling between TO oscillators ($\omega'$=0, as sketched out) is given in Fig. 2a.

Such contour modeling of the TO $\{\varepsilon_r(\omega,x) \to \infty\}$ $Cd_{1-x}Be_xTe$ Raman spectra in their $x$-dependence is achieved within a four-mode $\{2 \times (Cd-Te), 2 \times (Be-Te)\}$ version of the percolation model by calculating $Im\{\varepsilon_r(\omega,x)\}$ – in a crude approximation[S6,63] (see main text). At 0 GPa, the non-polar TO's are presumably decoupled and hence are explicitly assigned by specifying both the bond vibration and the local environment (via a subscript and a superscript, respectively), i.e., $\{TO_{Cd-Te}^{Be}, TO_{Cd-Te}^{Cd}, TO_{Be-Te}^{Be}, TO_{Be-Te}^{Cd}\}$. A classical form is used for $\varepsilon_r(\omega,x)$ including a linear background electronic contribution $\varepsilon_\infty(x)$ at high (visible) frequencies besides the phonon (far-infrared) one, i.e., four oscillators in total ($p$=1 to 4) modeled as $p$-Lorentzian functions. In each Lorentzian, the numerator represents the available amount of oscillator strength per $p$-mode $S_p^0(x)$ that scales as the parent value ($S_p^0 = \varepsilon_{\infty,p} \cdot \Omega_p^2/\omega_{T,p}^2$, with $\Omega_p^2 = \omega_{L,p}^2 - \omega_{T,p}^2$) weighted by the $p$-oscillator fraction ($f_p$-term below), directly impacting the $p$-type TO Raman intensity. The denominator $(\omega_{T,p}^2(x) - \omega^2 - j\gamma_p\omega)$ monitors the position of the TO $p$-resonance in its $x$-dependence, given by $\omega_{T,p}^2(x) = k_p(x)/\mu_p$ (the numerator and denominator referring to the effective mechanical bond force constant of oscillator-$p$ and to the reduced atomic mass of the $p$-type chemical bond, respectively), with the phonon damping $\gamma_p$ (introducing a friction force) fixing the full width at half maximum of the $p$-type Raman peaks (taken minimal in Fig. 2a, i.e., 1 cm$^{-1}$, for a clear resolution of neighboring features and direct comparison of the Raman instensities). We assume linear $k_p(x)$ vs. $x$ variations – in the spirit of the historical modified-random-element-isodisplacement (1-bond→1-mode) model[26] used to describe the Raman spectra of semiconductor alloys, leading to quadratic $\omega_{T,p}^2(x)$ variations. By analogy with $Zn_{1-x}Be_x$-chalcogenides[25,40], a sensitivity of Be-Te vibrations to crystalline environment limited to nearest neighbors is considered. The corresponding 1D-oscillators behind the four TO's (specified in brackets above) can be casted as $\{Te(Cd-Te)Be, Te(Cd-$ $Te)Cd, Te(Be − Te)Be, Te(Be − Te)Cd$}, participating with weights {$f_{Cd-Te}^{Be} = x \cdot (1 − x)$, $f_{Cd-Te}^{Cd} = (1 − x)^2$, $f_{Be-Te}^{Be} = x^2$, $f_{Be-Te}^{Cd} = x \cdot (1 − x)$}. The TO Raman intensities (color code in Fig. 2a) scale accordingly. Hence the two TO submodes forming a given (Cd-Te or Be-Te) doublet exhibit comparable Raman intensities at $x$~0.5.

Altogether, this leaves ten input parameters in total, out of which six relate to the parent compounds ($\varepsilon_{\infty,p}, \Omega_p, \omega_{T,p}$) and four to the alloy taken in its (Cd,Be)-dilute limits. All parameters are determined *ab initio*, without leaving any adjustable one. The input frequencies (two per TO branch) are derived by implementing a simple *ab initio* (AIMPRO) protocol onto impurity Cd-duo and Zn-duo motifs (referring to the $i$, $ii$ and $iv$ end frequencies in Fig. 2a), the parent $\omega_{T,p}$ (bulk TO) frequencies (symbolized $iii$ in Fig. 2a) coming out as by-products. The parent TO-LO splittings together with the parent $\varepsilon_{\infty,p}$ values, needed to estimate the parent oscillator strengths – see methods, are obtained by resorting to QE.

Fair phenomenological modeling of the pressure-induced interplay between Raman intensities is achieved within a linear dielectric approach of the relaxed-inverted Be-Te percolation doublet of Cd$_{1-x}$Be$_x$Te at $x$~1 (Fig. 2f) – apparent in *ab initio* data (Fig. 2e), by taking into account a (weak) mechanical coupling ($\omega'$~50 cm$^{-1}$ – see main text). A relevant expression for the corresponding Raman cross section has lately been derived – see Ref.[40] and specifically Eq. (7) therein. A simplified form is currently used in which the coupling term at the numerator is disregarded compared with the two main terms standing for the raw-uncoupled oscillators. The analysis is further displaced from the Be-dilute limit ($x$~1) to minor Be contents ($x$~0.8 in this case) for more convenience (see main text) – while artificially preserving the peak frequencies as such. A minimal phonon damping is uniformly taken (1 cm$^{-1}$) for clarity.

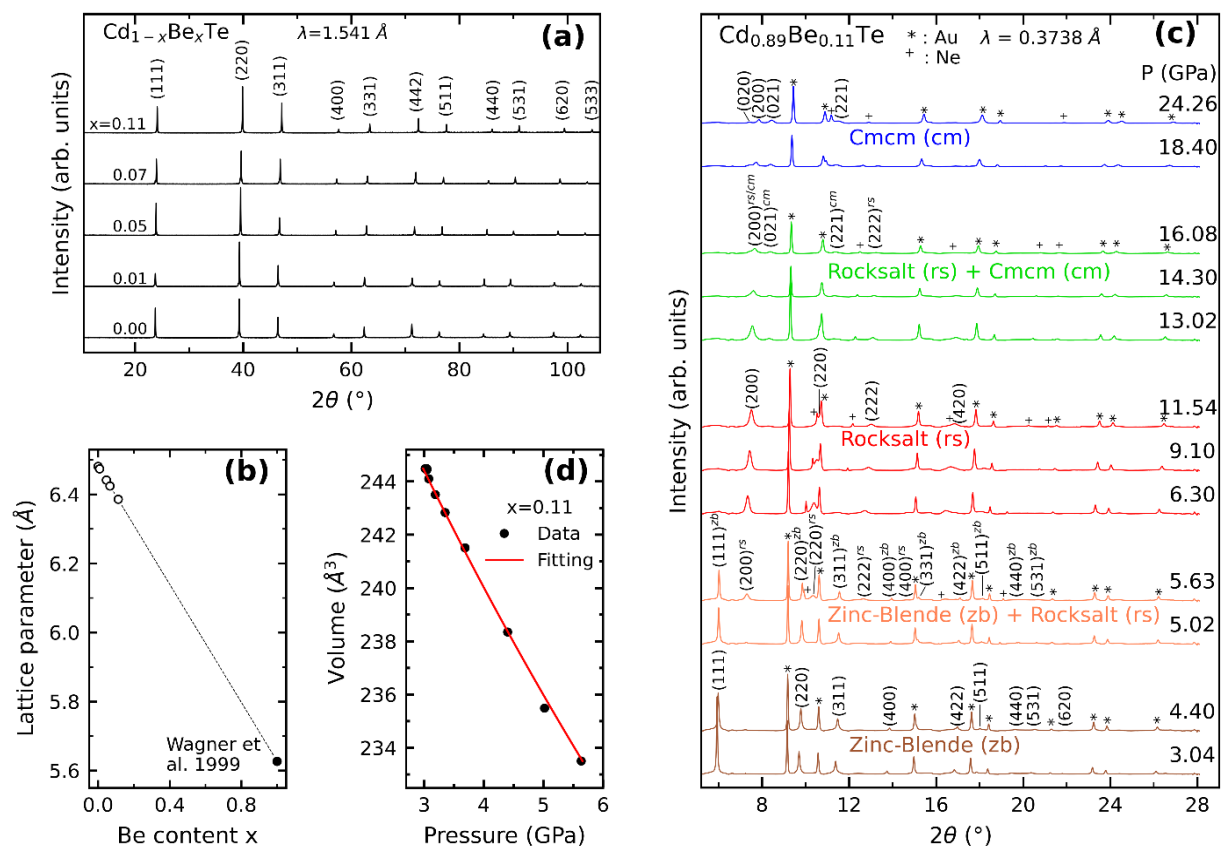

**Figure S1** | **(High-pressure) $Cd_{1-x}Be_xTe$ ($x \leq 0.11$) X-ray diffraction data. (a)** Powder $Cd_{1-x}Be_xTe$ X-ray diffractograms obtained at ambient pressure in laboratory. **(b)** Corresponding $x$-dependence of the lattice parameter. The BeTe value, taken from the literature[47], is added to complete the trend. The linearity is emphasized (dotted line). **(c)** selection of $Cd_{0.89}Be_{0.11}Te$-powder X-ray diffractograms obtained at increasing pressure. The individual peaks are labelled via the (hkl) Miller indices of the corresponding diffraction planes in various (zincblende-zb, rocksalt-rs, Cmcm-cm) structural phases. Additional diffraction peaks originate from Au and Ne used for pressure calibration and as the pressure transmitting medium, respectively. **(d)** Corresponding pressure dependence of the unit cell volume fitted to the Birch-Murnaghan equation of state.

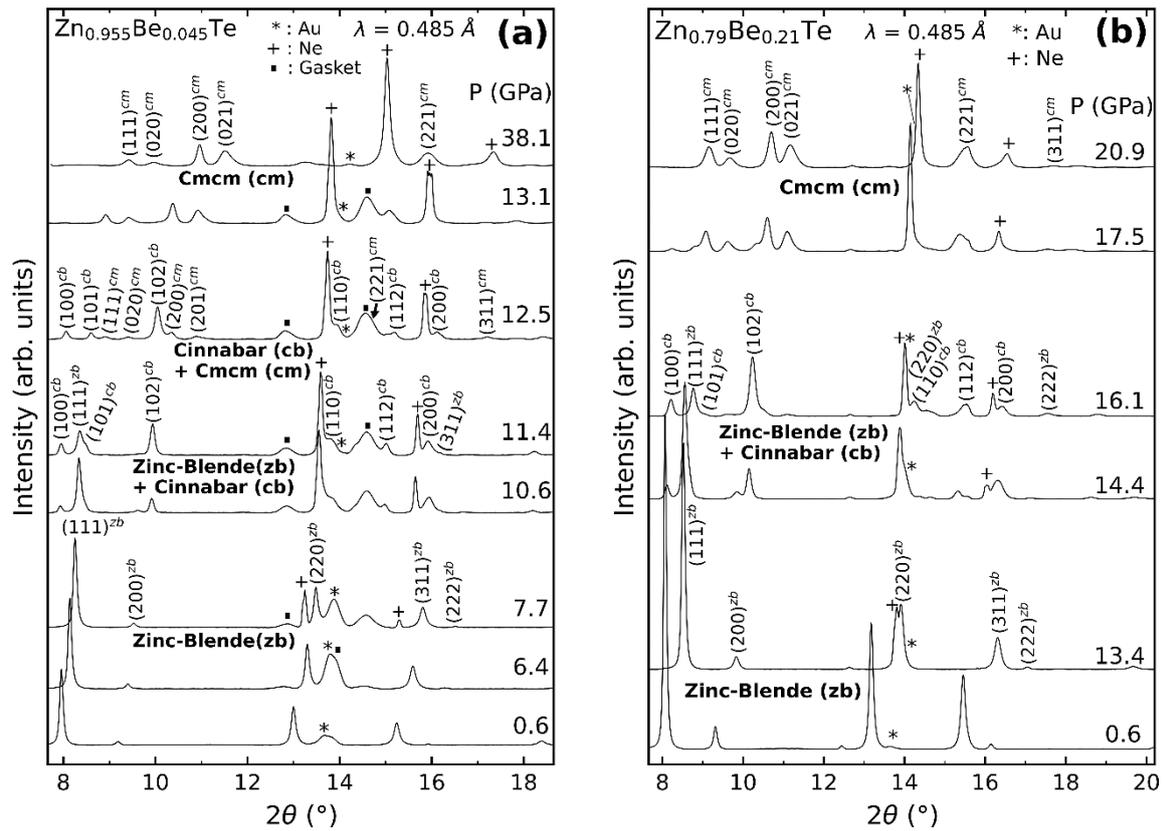

**Figure S2 | High-pressure $Zn_{1-x}Be_xTe$ X-ray diffraction data at moderate Be content.** Selection of **(a)** $Zn_{0.955}Be_{0.045}Te$ and **(b)** $Zn_{0.79}Be_{0.21}Te$ powder X-ray diffractograms obtained at increasing pressure. The individual peaks are labelled via the (hkl) Miller indices of the corresponding diffraction planes in various (zincblende-zb, rocksalt-rs, Cmcm-cm) structural phases. Additional diffraction peaks originate from Au and Ne used for pressure calibration and as the pressure transmitting medium, respectively, and from the gasket, as specified.

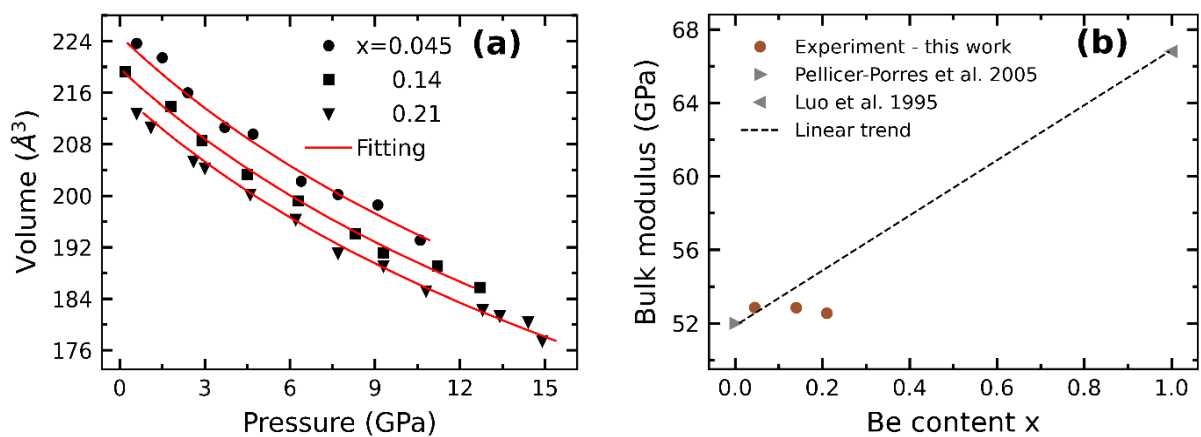

**Figure S3 | Zn$_{1-x}$Be$_x$Te bulk modulus. (a)** Pressure dependencies of the unit cell volume – as derived from the high-pressure Zn$_{1-x}$Be$_x$Te X-ray diffractograms (partially reported in Fig. S2) – fitted to the Birch-Murnaghan equation of state, including similar data obtained with Zn$_{0.86}$Be$_{0.14}$Te in the same run of experiment. **(b)** Resulting $B_0$ vs. $x$ variations.

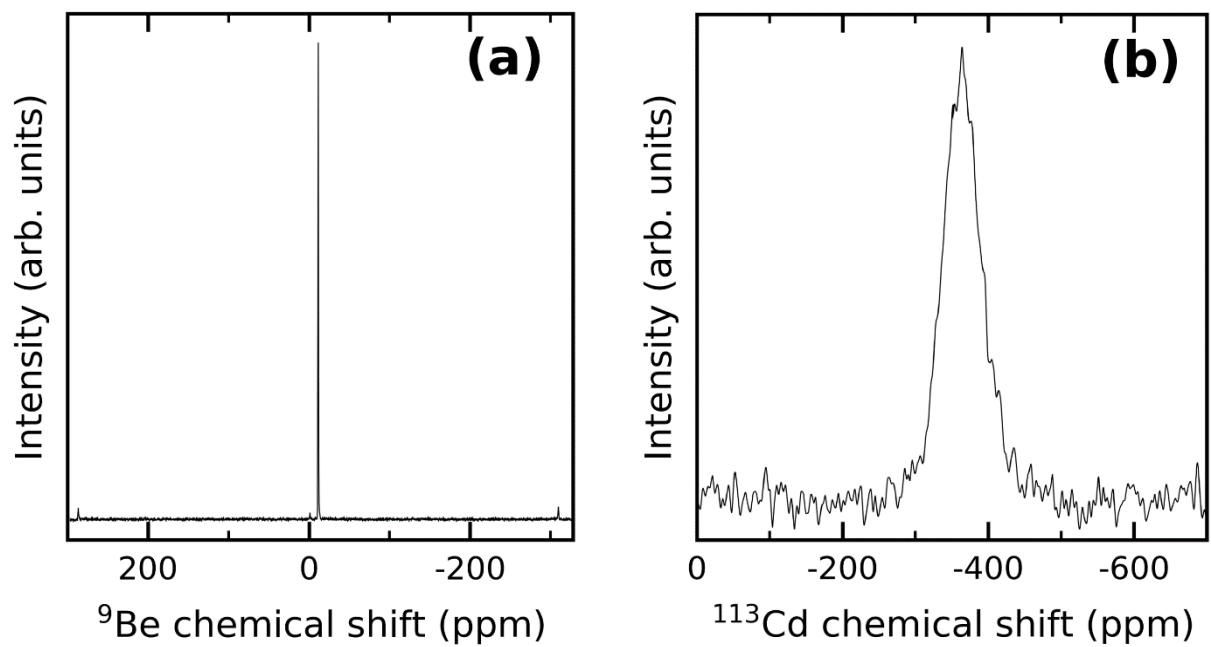

**Figure S4 | Substituent-related $Cd_{0.93}Be_{0.07}Te$ NMR spectra. (a)** $^9Be$ and **(b)** $^{113}Cd$ NMR standard direct acquisition spectra.

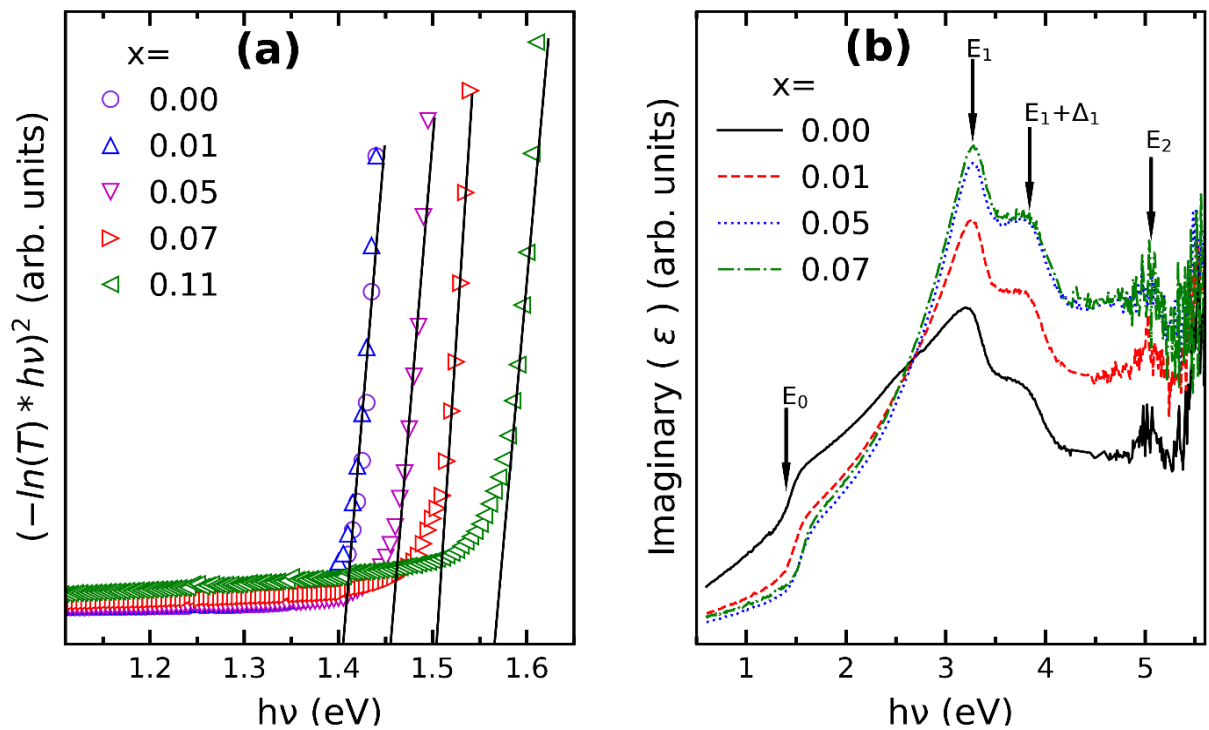

**Figure S5 | Cd$_{1-x}$Be$_x$Te transmission and ellipsometry data. (a)** Cd$_{1-x}$Be$_x$Te Tauc plot (curves) of transmission data (symbols) giving access to $E_0$. **(b)** Corresponding ellipsometry data obtained by direct (model free) wavelength-per-wavelength inversion of the sine and cosine of the depolarization angles measured by ellipsometry. The main electronic transitions are indicated.

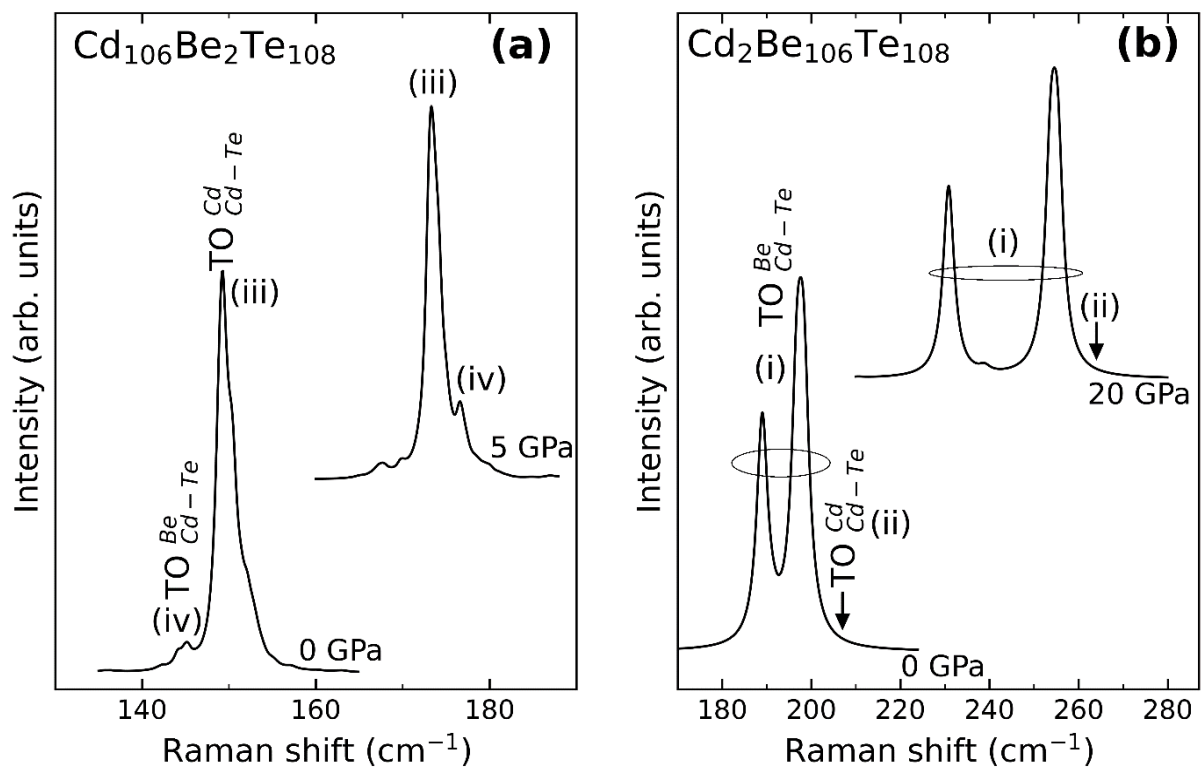

**Figure S6 | High-pressure *ab initio* (AIMPRO) CdTe-like Raman signals of $Cd_{1-x}Be_xTe$ generated by impurity-duos ($x\sim 0,1$). (a)** $Cd_{106}Be_2Te_{108}$. **(b)** $Cd_2Be_{106}Te_{108}$. Under pressure the CdTe-like branches either cross (as emphasized by a curved arrow) or freeze (the concerned submode is spotted by vertical arrows) at the resonance, as sketched out.

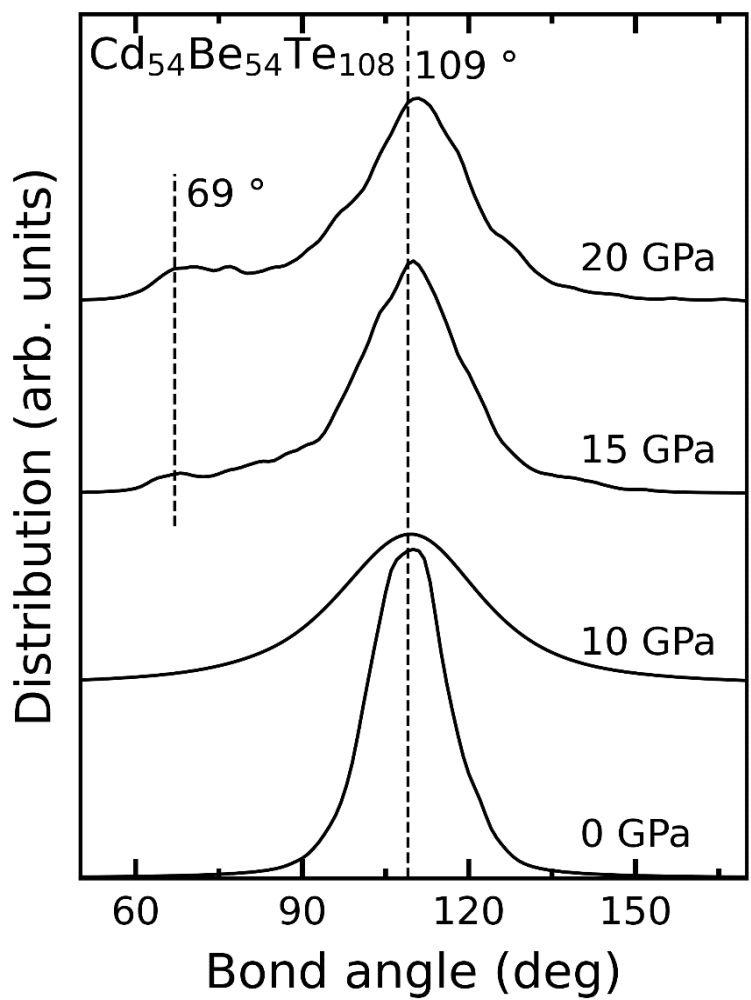

**Figure S7 | (High-pressure)** *ab initio* **(AIMPRO) insight into the $Cd_{54}Be_{54}Te_{108}$ lattice relaxation.** Pressure dependence of the *ab initio* bond-angle distribution within the large (216-atom) $Cd_{54}Be_{54}Te_{108}$ zincblende-type supercell optimized to a random Ce↔Be substitution bonds used to calculate the *ab initio* (AIMPRO) Raman spectra discussed in the main text (Fig. 2d). The nominal angle value in the zincblende structure (109°, dotted line) is indicated, for reference purpose. A significant deviation from 15 GPa onwards is emphasized (69°, dotted line).

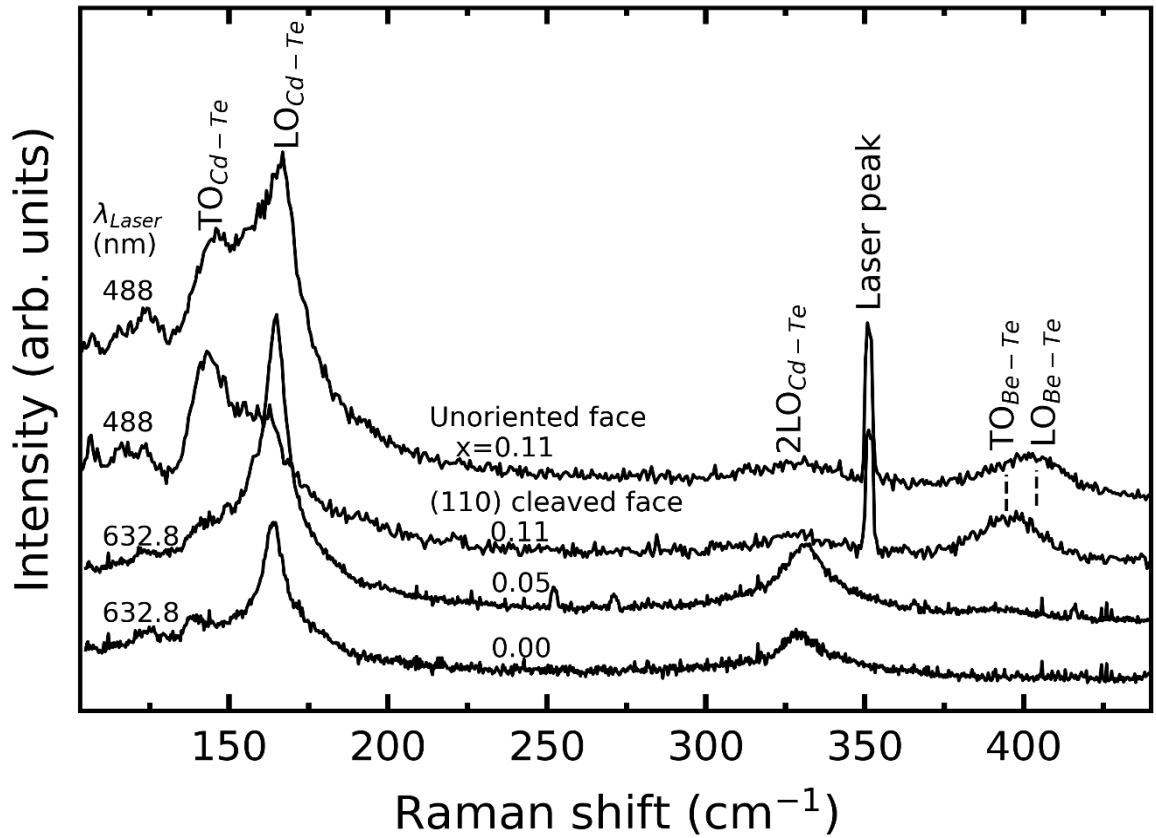

**Figure S7 | Experimental Cd$_{1-x}$Be$_x$Te ($x \leq 0.11$) Raman spectra at ambient conditions.** LO-like Cd$_{1-x}$Be$_x$Te Raman spectra taken in the backscattering geometry on unoriented crystal faces polished to optical quality in near-resonant conditions with the red (632.8 nm) and blue (488.0 nm) laser lines (Fig. 1b). A pure-TO Cd$_{0.89}$Be$_{0.11}$Te Raman spectrum taken with the 488.0 nm laser line in the backscattering geometry at normal incidence onto a (110)-cleaved face is added, for comparison. Paired dotted lines mark a finite TO-LO splitting at $x$=0.11.